\documentclass[12pt,titlepage]{utarticle}

\usepackage{amsmath,amssymb}
\usepackage{graphicx}

\newcommand{\be}{\begin{eqnarray}}
\newcommand{\non}{\nonumber \\}
\newcommand{\V}{\mathcal{V}}

\newcommand{\ee}{\end{eqnarray}}
\newcommand{\bdm}{\begin{displaymath}}
\newcommand{\edm}{\end{displaymath}}

\newcommand{\ba}{\begin{array}}
\newcommand{\ea}{\end{array}}
\newcommand{\Mpl}{M_{\mathrm{Pl}}}

\numberwithin{equation}{section}

\newcommand{\cm}{c^{(-)}_k}
\newcommand{\kax}{k_{\mathrm{max}}}
\newcommand{\kin}{k_{\mathrm{min}}}

\begin{document}

\baselineskip=15.5pt 

\title{Resonant Non-Gaussianity}

\author{Raphael Flauger
   \address{
      Department of Physics,\\
      Yale University,\\
      New Haven, CT 06520, USA\\
   }
       and Enrico Pajer
   \address{
      Department of Physics,\\ 
      Cornell University,\\
      Ithaca, NY 14853, USA\\
      {~}\\[.5cm]
      {\rm {\it e}-mail}\\[.1cm]
      \emailt{raphael.flauger@yale.edu}\\
      \emailt{ep295@cornell.edu}\\
   }
}
\Abstract{We provide a derivation from first principles of the primordial bispectrum of scalar perturbations produced during inflation driven by a canonically normalized scalar field whose potential exhibits small sinusoidal modulations. A potential of this type has been derived in a class of string theory models of inflation based on axion monodromy. We use this model as a concrete example, but we present our derivations and results for a general slow-roll potential with superimposed modulations. We show analytically that a resonance between the oscillations of the background and the oscillations of the fluctuations is responsible for the production of an observably large non-Gaussian signal. We provide an explicit expression for the shape of this \textit{resonant non-Gaussianity}. We show that there is essentially no overlap between this shape and the local, equilateral, and orthogonal shapes, and we stress that resonant non-Gaussianity is \textit{not} captured by the simplest version of the effective field theory of inflation. We hope our analytic expression will be useful to further observationally constrain this class of models.}
\maketitle
\tableofcontents

%%%%%%%%%%%%%%%%%%%%%%%%%%%%%%%%%%%%%%%%%%%%%%%%%%%%%%%%%%%%%%%%%%%%%%%%%%%%%%%%%%%%%%%%%%%%%%%%%%%%%%%%%%%%%

\section{Introduction and Summary}

The study of anisotropies in the cosmic microwave background radiation over the past two decades has dramatically improved our understanding of the early universe. 
There is now strong evidence that the anisotropies we see today originated from primordial fluctuations generated in the very early universe, and we have learned that these primordial fluctuations have a nearly scale-invariant spectrum. Furthermore, the data still contains no evidence for a deviation from adiabaticity, or Gaussianity~\cite{Komatsu:2010fb}. Even though the case is by no means closed, these properties of the primordial fluctuations certainly support the idea that they originated as quantum fluctuations during inflation, a phase of nearly exponential expansion of the universe~\cite{Guth:1980zm}.

While observations are now good enough to rule out some of the simplest inflationary models involving only a single slowly rolling field with canonical kinetic term, other models in this class are still compatible with all existing data. These models predict an adiabatic spectrum with primordial non-Gaussianities that are too small to be observed, but this is not a generic prediction of inflation. Many models of inflation have been constructed and studied that can lead to an observable departure from Gaussianity. Some popular possibilities are models with multiple fields, non-canonical kinetic terms, light spectator fields and a violation of slow-roll. Beyond an existence proof that observably large non-Gaussianities can be generated, these models provide us with useful theoretical expectations to guide our search.

For a Gaussian signal, all odd $n$-point functions vanish and the higher even $n$-point functions are given in terms of sums of products of the two-point function. The most straightforward way to look for a departure from Gaussianity is then to look for a non-zero three-point function. In Fourier space the three-point function depends on three momenta. Translational invariance of the background geometry ensures that these momenta add up to zero and thus form a triangle. Rotational invariance furthermore dictates that the three-point function can only depend on the three independent scalar products of these momenta. 
The information contained in the three-point function can thus be captured by a function of three variables, that can be thought of as two angles and one side of the triangle.
Since the dependence is a priori completely arbitrary, a model independent measurement would be desirable and would provide a precious criterion to discriminate between otherwise indistinguishable models. Unfortunately, progress in this direction is very hard. (For a review see {\it e.g.}~\cite{Liguori:2010hx}). Essentially all phenomenological analyses start from some explicit form of the three-point function  guided both by theoretical expectations and by the simplicity of the numerical analysis necessary to compare it with the data. (See, however,~\cite{Fergusson:2009nv}.) Once a ``shape'' has been chosen, only the amplitude of this type of non-Gaussianity remains as a parameter, which is conventionally called $f^\text{shape}$. So far only a handful of scale-invariant shapes have been looked for in the data. 
The most recent observational bounds on the magnitude for various shapes from the 7-year WMAP data at 95\% CL are~\cite{Komatsu:2010fb}:
\begin{center}
\begin{tabular}{lc}
local non-Gaussianity & $-10<f^\text{local}<74$\,,\\ 
equilateral non-Gaussianity& $-214<f^\text{equil}<266$\,,\\ 
orthogonal non-Gaussianity& $-410<f^\text{ortho}<6$\,.\\
\end{tabular}
\end{center}
Planck data will make it possible to tighten the error bars by about a factor of five, and we may soon find out whether it is necessary to go beyond the simplest models of inflation. 

As already briefly mentioned, departures from the slow-roll condition can potentially lead to large non-Gaussianities. 
Two conceptually distinct possibilities were first explored by Chen, Easther, and Lim. An otherwise smooth potential might either exhibit a sharp, localized feature~\cite{Chen:2006xjb} (see also~\cite{Bean:2008na}), or it might display a periodic modulation that averages to zero~\cite{Chen:2008wn}. 
Chen, Easther and Lim have performed a numerical analysis of both scenarios~\cite{Chen:2006xjb,Chen:2008wn}. They show that a large non-Gaussian signal can be produced without violating the constraints on these models from measurements of the two-point function. They also provide a heuristic estimate of the signal for equilateral configurations for the case of resonant production.

In this work, we analytically compute the scalar primordial bispectrum generated for a modulated potential for arbitrary momentum configurations from first principles. Our work is motivated by a class of models derived from string theory~\cite{McAllister:2008hb, Flauger:2009ab}, but let us stress that these are not the only models in which such oscillations are expected to arise. In large field inflation, the inflaton potential must be flat over a range in field space large compared to $\Mpl$. From the point of view of effective field theory, this seems unnatural unless there is an underlying shift symmetry. Axions are thus natural candidates for the inflaton in large field inflation. It then seems plausible that the potential might receive small periodic contributions from non-perturbative effects. These periodic contributions might be due to instantons in a gauge sector the axion couples to, or, in the context of string theory, they might arise from Euclidean branes or world-sheet instantons. Whether string inspired or not, as soon as we invoke the shift symmetry of axions to explain why the inflaton potential is so flat, we should admit the possibility of small periodic modulations in the potential which may lead to  observational consequences. To be specific, the potentials we will consider are of the form 
\be \label{V}
V(\phi)=V_0(\phi)+\Lambda^4\cos \left(\frac{\phi}{f}\right)\,,
\ee
where $\phi$ is a canonically normalized real scalar field, and $V_0(\phi)$ is assumed to admit slow-roll inflation in the absence of modulations, ${\it i.e.}$ for $\Lambda=0$. 
A particular model of this form with $V_0(\phi)=\mu^3\phi$ was obtained from a string theory construction in~\cite{McAllister:2008hb, Flauger:2009ab}, and we will sometimes focus on this special case for concreteness.
The parameters $\Lambda$ and $f$ have dimensions of a mass and are a priori undetermined. However, consistency of a more fundamental description of the system, in our case string theory, will typically limit them to lie in a certain range.  
We will assume that the potential is monotonic at least near the values of the scalar field around which the modes we observe in the cosmic microwave background (CMB) exit the horizon and do not consider models in which the inflaton gets trapped. This can be summarized by requiring that the monotonicity parameter 
\be
b_*\equiv\frac{\Lambda^4}{V_0'(\phi_*)f}<1\,.
\ee
Except for a linear potential, this parameter depends on the value of the field. We will evaluate it at $\phi=\phi_*$, the value of the scalar field at the time when the mode with comoving momentum equal to the pivot scale $k=k_*$ exits the horizon. 
To be compatible with WMAP data, the monotonicity parameter must satisfy $b_*\ll1$ for both linear~\cite{Flauger:2009ab} and quadratic $V_0(\phi)$~\cite{Pahud:2008ae}. Other potentials of this form have not been compared to the data, but we expect this to be true for the general case and treat $b_*$ as an expansion parameter. 

Our main result is that the three-point function of scalar curvature perturbations to linear order in $b_*$ at some late time $t$, when the modes have exited the horizon, takes the form\footnote{We have set $\Mpl=1$ in this formula and will do so throughout the paper. As usual, it can be re-inserted by dimensional analysis.}
\begin{multline}\label{resu}
\langle\mathcal{R}({\bf k_1},t)\mathcal{R}({\bf k_2},t)\mathcal{R}({\bf k_3},t)\rangle=(2\pi)^7\Delta_\mathcal{R}^4\frac{1}{k_1^2k_2^2k_3^2} \delta^3({\bf k_1}+{\bf   k_2}+{\bf k_3})\\\times f^\text{res}\left[\sin\left(\frac{\sqrt{2\epsilon_*}}{f}\ln K/k_*\right)+\frac{f}{\sqrt{2\epsilon_*}} \sum\limits_{i\neq j}                           \frac{k_i}{                   k_j}\cos\left(\frac{\sqrt{2\epsilon_*}}{f}\ln K/k_* \right)+\dots\right]\,.
\end{multline}
with
\begin{equation}\label{fres}
f^\text{res}=\frac{3 b_*\sqrt{2\pi}}{8}\left(\frac{\sqrt{2\epsilon_*}}{f}\right)^{3/2}\,,
\end{equation}
where $\epsilon_*$ denotes the value of the slow-roll parameter derived from the smooth part of the potential $V_0(\phi)$, evaluated at the time the pivot scale $k_*$ exits the horizon. The quantities $\phi_*$ and $\epsilon_*$ are model dependent, but are easy to calculate for any given model. The comoving momentum $K=k_1+k_2+k_3$ is the perimeter of the triangle in momentum space.
The dots in \eqref{resu} stand for terms that can be neglected either because they are suppressed by higher powers in the slow-roll parameters for the smooth part of the potential or by positive powers of $f/\sqrt{2\epsilon_*}$. From \eqref{fres} one can see that large non-Gaussianity requires $f/\sqrt{2\epsilon_*}\ll1$. So this will be the regime of interest in this the paper. The second term is suppressed compared to the first by a factor of $f/\sqrt{2\epsilon_*}$. It is negligible except for squeezed triangles where one of the momenta is much less than the other two. It ensures that the consistency relation of~\cite{Maldacena:2002vr} (see also \cite{Creminelli:2004yq}) holds. 

We compare our analytic result with the numerical analysis of~\cite{Hannestad:2009yx} for the linear potential of axion monodromy. We find agreement with their numerical results for the bispectrum at the per cent level. 
We compare our results with the numerical analysis of \cite{Chen:2008wn} for the quadratic potential. Identifying their parameter $\tilde{P}^2$ with $9\Delta_\mathcal{R}^4(k_*)/10$, we find agreement with their numerical results as well.

This type of non-Gaussianity, which we refer to as \textit{resonant non-Gaussianity} following the nomenclature in~\cite{Chen:2008wn}, is nearly orthogonal to all commonly studied shapes. The cosine defined in~\cite{Babich:2004gb,Fergusson:2008ra} is less than $10\%$ for the entire range of parameters relevant for axion monodromy inflation.
This is intuitively clear. The shape of resonant non-Gaussianity rapidly oscillates around zero while the other shapes are slowly varying. As the number of oscillations increases, the cosine decreases. The number of oscillations over the scales observed in the cosmic microwave background is approximately given by $\sqrt{2\epsilon_*}/f$. For our string theory example, this implies that an axion decay constant $f=10^{-3} $ leads to about $90$ periods in the cosmic microwave background. The cosine with local, equilateral, and orthogonal shapes is then around $2\%$, and the largest amount of non-Gaussianity for this value of $f$ that is consistent with the observational constraints on the two-point function is $f^\text{res}\approx 80$.  

As we will explain, the resonant effect we are studying arises from a term in the interaction Hamiltonian that is higher order in the slow-roll parameters. As a consequence, it is not captured by the simplest version of the effective field theory of inflation~\cite{Cheung:2007st} (see also~\cite{Weinberg:2008hq}). A more detailed discussion is postponed to Section~\ref{s:disc}.

The paper is organized as follows. In Section~\ref{s:power}, we derive the time evolution of the curvature perturbation for a potential~\eqref{V} and obtain the power spectrum. The derivation is independent of the ones presented in \cite{Flauger:2009ab}. It agrees with the results that were found there for the linear potential. In Section~\ref{s:bis}, we calculate the bispectrum and check that it fulfills the consistency relation of~\cite{Maldacena:2002vr} for squeezed triangles. In Section~\ref{s:cosine}, we compute the overlap between the shape of resonant non-Gaussianity and the three most common shapes in the literature that have been compared with the data. We show that it is very small for the range of parameters relevant for our string theory example. Section~\ref{s:disc} contains a discussion of our results. In Appendix \ref{a:gen}, we present the details of the calculation of the inflationary background solutions for the potential \eqref{V}. Appendix~\ref{a:sr} provides the relation between various different definitions for slow-roll parameters commonly used in the literature.

%%%%%%%%%%%%%%%%%%%%%%%%%%%%%%%%%%%%%%%%%%%%%%%%%%%%%%%%%%%%%%%%%%%%%%%%%%%%%%%%%%%%%%%%%%%%%%%%%%%%%%%%%%%%%

\section{The Mode Functions and the Power Spectrum} \label{s:power}

As we show in Appendix~\ref{a:gen}, the background solution for a scalar field with potential~\eqref{V} can be derived to first order in $b_*$ and to leading order in the slow-roll parameters of $V_0$. In the limit $f\ll\sqrt{2\epsilon_*}$, which, as we shall see, is the regime where large non-Gaussianities are generated, the solution is well approximated by
\be
\phi(t)=\phi_0(t)-\frac{3 b_* f^2}{\sqrt{2\epsilon_*              }}\sin\left(\frac{\phi_0(t)}{f}\right)\,.
\ee
Here $\phi_*$ is the value of the scalar field when the pivot scale $k=k_*$ exits the horizon,\footnote{For numerical calculations, we will take $k_*=0.002\,\text{Mpc}^{-1}$. The value of $\phi_*$ is model dependent. For the linear potential we use $\phi_*=10.88$.} $\epsilon_*$ is the slow-roll parameter in the absence of modulations evaluated at $\phi_0=\phi_*$, and $\phi_0$ is the solution for the scalar field in the absence of modulations, {\it i.e.} for $b_*=0$. 
It can be obtained as a function of time by integrating its equation of motion, but we will not need the result at this time. 

Since we will need them later, let us give the expressions for the Hubble slow-roll parameters\footnote{For the convenience of the reader we have collected in other possible definitions of the slow-roll parameters and formulae for the conversion in Appendix \ref{a:sr}.} 
\be\label{eq:Hsr}
\epsilon\equiv-\frac{\dot{H}}{H^2}\hskip 1cm \text{and}\hskip 1cm\delta=\frac{\ddot{H}}{2\dot{H}H}\,.
\ee
For the potential~\eqref{V}, it is convenient to calculate them in an expansion in the parameter $b_*$
\be
\epsilon=\epsilon_0 + \epsilon_1 + {\cal{O}}(b_*^2)\,,\\ 
\delta=\delta_0 + \delta_1 + {\cal{O}}(b_*^2)\,.
\ee
In the slow-roll approximation for $\phi_0$, and for $f\ll\sqrt{2\epsilon_*}$, one finds
\be
&&\epsilon_0=\epsilon_*,\;\;\;\;\delta_0=\epsilon_*-\eta_*\,,\\
&&\epsilon_1=-3b_*f\sqrt{2\epsilon_*}\cos\left(\frac{\phi_0(t)}{f}\right) \,,\\
&&\delta_1=-3b_*\sin\left(\frac{\phi_0(t)}{f}\right)\label{eq:d1}\,,
\ee 
where $\eta_*$ and $\epsilon_*$ are the values of the potential slow-roll parameters $\eta_{V_0}\equiv V_0''/V_0$ and $\epsilon_{V_0}\equiv(V_0'/V_0)^2/2$ derived from the smooth part of the potential $V_0$, evaluated at the time the pivot scale $k_*$ exits the horizon. Notice that in this regime both $\epsilon_1\ll1$ and $\delta_1\ll1$ as long as $b_*\ll1$. On the other hand this is not the case for higher slow-roll parameters. For instance, one has
\be
\dot\delta_1/H=3b_* \frac{\sqrt{2\epsilon_*}}{f}                        \cos\left(\frac{\phi_0(t)}{f}\right)\label{eq:dd1}\,,
\ee
which becomes large for small $f/\sqrt{2\epsilon_*}$.

Let us now turn to the spectrum of scalar perturbations. As in~\cite{Flauger:2009ab}, we will choose a slicing such that $\delta\phi({\bf x},t)=0$, and use a spatial diffeomorphism to bring the scalar perturbations in the spatial part of the metric into the form
\be
\delta g_{ij}({\bf x},t)=2a(t)^2\mathcal{R}({\bf x},t)\delta_{ij}\,.
\ee
The translational invariance of the background makes it convenient to look for the solution as a superposition of Fourier modes
\be\label{eq:rx}
\mathcal{R}({\bf x},t)=\int\frac{d^3{\bf k}}{(2\pi)^{3}}\mathcal{R}({\bf k},t)e^{i{\bf k}\cdot{\bf        x}}\,,
\ee
where ${\bf x}$ are the comoving coordinates and ${\bf k}$ denotes the comoving momentum of the mode.
Rotational invariance together with reality (or Hermiticity) of $\mathcal{R}({\bf x},t)$ imply that the Fourier components of the most general solution can be written in the form
\be\label{eq:rk}
\mathcal{R}({\bf k},t)=\mathcal{R}_k(t)a({\bf k})+\mathcal{R}^*_k(t) a^\dagger(-{\bf k})\,,
\ee
where $k$ is the magnitude of the comoving momentum ${\bf k}$, $a({\bf k})$ can be thought of as a stochastic parameter in the classical theory or as an annihilation operator in the quantum theory, and $\mathcal{R}_k(t)$ is the mode function. When thought of as creation and annihilation operators $a^\dagger({\bf k}')$ and $a({\bf k})$ satisfy the commutation relation
\be
\left[ a({\bf k}), a^\dagger({\bf k}')\right]=(2\pi)^3\delta^3({\bf k}-{\bf k'})\,.
\ee
The time evolution of the mode function $\mathcal{R}_k(t)$ is governed by the Mukhanov-Sasaki equation~\cite{Mukhanov:1985rz,Sasaki:1986hm}. For small $\epsilon$, it can be written in the form~\cite{Weinberg:2008zzc}
\be \label{MSx}
\frac{d^2\mathcal{R}_k}{dx^2}-\frac{2(1+2\epsilon+\delta)}{x}\frac{d\mathcal{R}_k}{dx}+\mathcal{R}_k=0\,.
\ee
The initial conditions are such that for $x\gg 1$
\be\label{in cond}
\mathcal{R}_k(x)\to-\frac{H}{\sqrt{2k}a\dot\phi}e^{ix}\,,
\ee
where we have used the notation $x\equiv-k\tau$ with conformal time $\tau$ defined as $\tau\equiv \int^t\frac{dt'}{a(t')}$. The Mukhanov-Sasaki equation implies that for $x\ll1$ the mode function $\mathcal{R}_k(x)$ approaches a constant which we denote by $\mathcal{R}_k^{(o)}$, where the superscript ${(o)}$ indicates that the mode is outside the horizon.
It is related to the quantity $\Delta_\mathcal{R}^2(k)$ that is commonly quoted to parameterize the primordial scalar power spectrum by
\be\label{eq:psr}
\left|\mathcal{R}_k^{(o)}\right|^2=2\pi^2 \frac{\Delta_\mathcal{R}^2(k)}{k^3}\,.
\ee
In the slow-roll approximation, {\it i.e.}~for $\epsilon\ll 1$, $\delta\ll 1$, and assuming $\dot\delta/H$ is small compared to both $\epsilon$ and $\delta$   
\be\label{eq:psd}
\Delta_\mathcal{R}^2(k)=\frac{H^2(t_k)}{8\pi^2\epsilon(t_k)} \quad \textrm{(slow-roll approximation)}\,,
\ee
where $t_k$ is the time at which the mode with comoving momentum $k$ exits the horizon.
However, as already pointed out in~\cite{Flauger:2009ab}, the slow-roll approximation breaks down in models with modulated potentials because the magnitude of $\dot\delta/H$ is no longer of quadratic order in $\epsilon$ and $\delta$. Furthermore, the slow-roll parameters are oscillatory functions whose frequency changes in time. When the frequency of this oscillation passes through twice the natural frequency of the mode, which is set by the momentum of the mode, parametric resonance occurs, which is not captured in the slow-roll approximation.

In~\cite{Flauger:2009ab} our main concern was the power spectrum so that only the asymptotic behavior of the mode function was needed but not its detailed behavior as a function of time. The main concern of this work is the calculation of the bispectrum for which the knowledge of the time dependence of the mode functions is important. So let us calculate it. We will neglect the effect of $\epsilon_0$ and $\delta_0$ in equation~\eqref{MSx} for simplicity, but they could be restored without too much extra trouble. Furthermore, we will make use of the fact that the amplitude of $\epsilon_1$ is suppressed compared to that of $\delta_1$ by a factor of $f\sqrt{2\epsilon_*}$ and drop it as well. The Mukhanov-Sasaki equation~\eqref{MSx} then becomes
\be 
\frac{d^2\mathcal{R}_k}{dx^2}-\frac{2(1+\delta_1(x))}{x}\frac{d\mathcal{R}_k}{dx}+\mathcal{R}_k=0\,.
\ee
As was shown in~\cite{Flauger:2009ab}, for the linear potential parametric resonance occurs around $x_\text{res}=1/(2f\phi_*)$. As we will see, for the general potential it happens at $x_\text{res}=\sqrt{2\epsilon_*}/(2f) $. For $x\gg x_\text{res}$, {\it i.e.}~much before the resonance occurred, we know that the effect of $\delta_1$ is negligible. Therefore the solution is\footnote{To keep the dependence of the mode function on the slow-roll parameters $\epsilon_0$ and $\delta_0$, one should replace $3/2$ by $3/2+2\epsilon_0+\delta_0$ in the discussion below.}
\be\label{eq:solunpert}
\mathcal{R}_k(x)=\mathcal{R}_{k,0}^{(o)}i\sqrt{\frac{\pi}{2}}x^{3/2}H_{3/2}^{(1)}(x)\,,
\ee
where $\mathcal{R}_{k,0}^{(o)}$ is the value of $\mathcal{R}_k(x)$ outside the horizon in the absence of modulations and is fixed by the initial condition \eqref{in cond}, and 
\be
i\sqrt{\frac{\pi}{2}}x^{3/2}H_{3/2}^{(1)}(x)=(1-i x)e^{i x}\,.
\ee
Similarly, for $x\ll x_\text{res}$, {\it i.e.}~long after the resonance has occurred, the background frequency is too high for the mode to keep up with it, and the effect of $\delta_1$ is again negligible. The solution there must take the form
\be
\mathcal{R}_k(x)&=&\mathcal{R}_{k,0}^{(o)}\left[c^{(+)}_k i\sqrt{\frac{\pi}{2}}x^{3/2}H_{3/2}^{(1)}(x)-c^{(-)}_k i\sqrt{\frac{\pi}{2}}x^{3/2}H_{3/2}^{(2)}(x)\right]\,.
\ee
A slight generalization of our derivation in~\cite{Flauger:2009ab} implies that at late times $c^{(+)}_k=1+\mathcal{O}(b_*^2)$.
It then seems natural to look for a solution of the form
\be\label{eq:solpert}
\mathcal{R}_k(x)=\mathcal{R}_{k,0}^{(o)}\left[i\sqrt{\frac{\pi}{2}}x^{3/2}H_{3/     2}^{(1)}(x)-c^{(-)}_k(x)i\sqrt{\frac{\pi}{2}}x^{3/2}H_{3/2}^{(2)}(x)\right]\,,
\ee
where $c^{(-)}_k(x)$ vanishes at early times and goes to $c^{(-)}_k$ at late times.
The Mukhanov-Sasaki equation then turns into an equation governing the time evolution of $c^{(-)}_k(x)$. To linear order in $b_*$, this equation is
\be\label{tt}
\frac{d}{dx}\left[e^{-2 i x}\left(1-\frac{i}{x}\right)\frac{d}{dx}\cm(x)\right]+e^{-2 i x}\frac{i}{x^2}\frac{d}{dx}\cm(x)=-2i \frac{\delta_1(x)}{x}\,.
\ee
For large $x$, which is where the resonance occurs as long as $f/\sqrt{2\epsilon_*}\ll1$, equation \eqref{tt} can be written in the form
\be
\frac{d}{dx}\left[e^{-2 i x}\frac{d}{dx}\cm(x)\right]=-2 i\frac{\delta_1(x)}{x}\,.
\ee
Using the expression for $\delta_1(x)$ given by equation~\eqref{eq:d1} together with\footnote{See Appendix~\ref{a:gen} for a details.}
\begin{equation}
\phi_0(x)=\phi_k+\sqrt{2\epsilon_*}\ln x\;\;\;\;\text{where} \;\;\;\;\phi_k=\phi_*-\sqrt{2\epsilon_*}\ln k/k_*\,,
\end{equation}
where $\phi_k$ is the value of the scalar field when the mode with comoving momentum $k$ exits the horizon,
this can immediately be integrated once to give
\be\label{eq:cmprime}
\frac{d}{dx}{c}^{(-)}_k(x)=-6ib\frac{f}{\sqrt{2\epsilon_*}}e^{2 i x}\cos\left(\frac{\phi_k}{f}+\frac{\sqrt{2\epsilon_*}}{f}\ln x\right)\,.
\ee
We will be able to use this result in the next section to argue that the modification of the mode function can be ignored when calculating the bispectrum to leading order in $b_*$. 
Let us now integrate this equation once again to get a better idea for what the function ${c}^{(-)}_k(x)$ looks like.
To separate the leading and subleading contributions, it is convenient to write $\cm(x)$ as
\begin{equation}\label{eq:cmeq}
{c}^{(-)}_k(x)=-3ib\frac{f}{\sqrt{2\epsilon_*}}\left[e^{-i\frac{\phi_k}{f}}\int\limits_\infty^x dx' \;e^{2 i x'-i\frac{\sqrt{2\epsilon_*}}{f}\ln x'}+e^{i\frac{\phi_k}{f}}\int\limits_\infty^x dx' \;e^{2 i x'+i\frac{\sqrt{2\epsilon_*}}{f}\ln x'}\right]\,.
\end{equation}
\begin{figure}[h!]
\begin{center}
\includegraphics[width=6.6in]{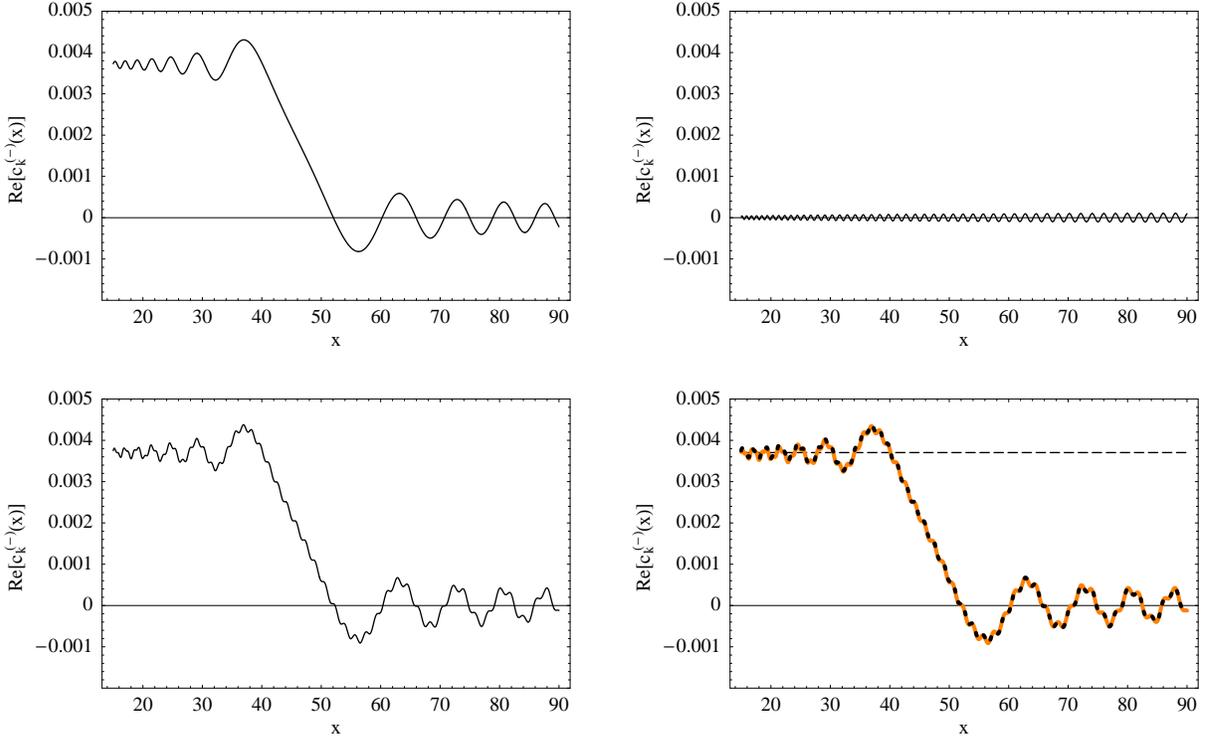}
        \caption{The top left shows the dominant contribution to $\text{Re}[{c}^{(-)}_k(x)]$ coming from the first integral in equation~\eqref{eq:cmeq}, while the top right shows the subdominant contribution to $\text{Re}[{c}^{(-)}_k(x)]$, which comes from the second integral in equation~\eqref{eq:cmeq}. The bottom left shows their superposition. The bottom right shows the superposition as black dotted line compared to the numerical solution in orange. As can be seen, the two results are essentially indistinguishable. The dashed line represents the asymptotic value for $x$ going to zero ({\it i.e.} outside the horizon) found by performing the integral in~\eqref{eq:cmeq2} in the stationary phase approximation. All plots are for a linear potential with $f=10^{-3}$, $b=10^{-2}$, $\phi_*=10.88$, and for a value of comoving momentum such that $\phi_k=10.7$. Notice that $\text{Re}[{c}^{(-)}_k(x)]$ changes around $x_\text{res}=1/2f\phi_*\approx 45$ as expected.}
      \label{fig:modefcn}
\end{center}
\end{figure}
After some manipulations, these integrals can be recognized as incomplete $\Gamma$-functions or closely related exponential integrals. 
The result can be written as 
\be \label{exact}
\cm(x)&=&-\frac32 b\frac{f}{\sqrt{2\epsilon_*}}\left\lbrace e^{\frac{\sqrt{2\epsilon_*}}{f}\left(\frac{\pi}{2}+i\ln2\right)}e^{-i\frac{\phi_k}{f}}\Gamma\left[1-i\frac{\sqrt{2\epsilon_*}}{f},-2ix\right]+       \right. \\&& \hspace{3cm} \left.+ e^{-\frac{\sqrt{2\epsilon_*}}{f}\left(\frac{\pi}{2}+i\ln2\right)}e^{i\frac{\phi_k}{f}}\Gamma\left[1+i\frac{\sqrt{2\epsilon_*}}{f},-2ix\right]                       \nonumber\right\rbrace \,.\ee
To gain some intuition, the stationary phase approximation is useful. The phase of the first integrand in \eqref{eq:cmeq} will become stationary near the resonance at $x=x_\text{res}$, while the phase of the second integrand is never stationary (since $x>0$). The first term on the right hand side of equations~\eqref{eq:cmeq} and \eqref{exact} will thus be the dominant contribution, with the subdominant contribution from the second term exponentially decreasing for decreasing $f/\sqrt{2\epsilon_*}$.
Figure~\ref{fig:modefcn} shows the leading and subleading contributions to the real part of ${c}^{(-)}_k(x)$, their superposition, as well as the comparison to the numerical result for the ${c}^{(-)}_k(x)$ from its evolution equation \eqref{tt} for a linear potential. Equation \eqref{tt} is valid for arbitrary $x$ and thus exact at linear order in $b_*$. 
As a byproduct, we have given an expression for $\cm\equiv\cm(0)$ of the form
\be\label{eq:cmeq2}
\cm=3ib\frac{f}{\sqrt{2\epsilon_*}}e^{-i\frac{\phi_k}{f}}\int\limits^\infty_0 dx \;e^{2 i x-i\frac{\sqrt{2\epsilon_*}}{f}\ln x}\,.
\ee
When evaluated exactly as in \eqref{exact} or using the stationary phase approximation, one finds that up to an unimportant phase 
\be \label{cm}
\cm=3b\sqrt{\frac{\pi}{2}}\left(\frac{f}{\sqrt{2\epsilon_*}}\right)^{1/2}e^{-i\frac{\phi_k}{f}}\,.
\ee
For the linear potential we can set $\sqrt{2\epsilon_*}=1/\phi_*$, and we find that this expression agrees with equation (3.40) in~\cite{Flauger:2009ab} for small $f\phi_*$.\footnote{This is true after dropping the same $k$-independent phase that was dropped there.}
Using equations~\eqref{eq:solpert} and \eqref{cm} as well as the behavior for the Hankel functions for small arguments, one finds that the primordial power spectrum of scalar fluctuations is of the form
\be
|\mathcal{R}_k^{(o)}|^2=|\mathcal{R}_{k,0}^{(o)}|^2\left[1+3b_*\left(\frac{ 2\pi f}{\sqrt{2\epsilon_*}}\right)^{1/2}\cos\left(\frac{\phi_k}{f}\right)\right]\,,
\ee
or equivalently
\be\label{eq:d2r}
\Delta_\mathcal{R}^2(k)=\Delta_\mathcal{R}^2(k_*)\left(\frac{k}{k_*}\right)^{n_s-1}\left[1+\delta n_s\cos\left(\frac{\phi_k}{f}\right)\right]\,\;\;\;\text{with}\;\;\;\;\delta n_s=3b_*\left( \frac{2\pi f}{\sqrt{2\epsilon_*}}\right)^{1/2}\,,
\ee
where once again $\phi_k$ is the value of the scalar field at which the mode with comoving momentum $k$ exits the horizon, $\epsilon_*$ is the value of $\epsilon_{V_0}$ when the pivot scale $k=k_*$ exits the horizon, and $b_*=\Lambda^4/V_0'(\phi_*)f$. We have restored the dependence on $\epsilon_0$ and $\delta_0$ through the appearance of the scalar spectral index $n_s$, which in the approximation we are using is given by $n_s=1-4\epsilon_0-2\delta_0=1-6\epsilon_*+2\eta_*$.

Everything we have said here about the primordial power spectrum for the scalar modes is valid for small $f/\sqrt{2\epsilon_*}$, which, as we will see, is the regime in which observable non-Gaussianity can be generated. In~\cite{Flauger:2009ab}, the interested reader can find the result for the linear potential for general $f\phi_*$.

We have performed numerical calculations to check these analytic results, and we find good agreement. Most of the numerical calculations were done for the linear potential relevant for axion monodromy inflation, but we have also performed some checks for the case of a quadratic potential as well as $V_0(\phi)=\mu^{10/3}\phi^{2/3}$ motivated by~\cite{Silverstein:2008sg}. Our numerical results for the amplitude of the modulations as well as the frequency agree with our analytic result at the per cent level in all cases.
The discrepancy between our analytic result and the numerical results for the power spectrum in~\cite{Hannestad:2009yx} can be traced to an initial value for $k/aH$ in their numerical calculation that was too small to capture the resonance for small axion decay constants. This issue will be easy to fix. 

%%%%%%%%%%%%%%%%%%%%%%%%%%%%%%%%%%%%%%%%%%%%%%%%%%%%%%%%%%%%%%%%%%%%%%%%%%%%%%%%%%%%%%%%%%%%%%%%%%%%%%%%%%%%%

\section{The Bispectrum}\label{s:bis}
Let us now turn to the calculation of the three-point function. To leading order in perturbation theory, the three-point function in the ``in-in'' formalism~\cite{Schwinger} (see also~\cite{Weinberg:2005vy,Adshead:2009cb} and references therein) is given by
\be
\langle\mathcal{R}({\bf k_1},t)\mathcal{R}({\bf k_2},t)\mathcal{R}({\bf k_3},t)\rangle=-i\int^t_{-\infty}\;dt'\langle\left[\mathcal{R}({\bf k_1},t)\mathcal{R}({\bf k_2},t)\mathcal{R}({\bf k_3},   t),H_I(t')\right]\rangle\,,
\ee
where the expectation value is taken in the in-vacuum, and the interaction Hamiltonian $H_I$ was first worked out in \cite{Maldacena:2002vr} (see also~\cite{Chen:2006nt,Seery:2005wm}). The term responsible for the dominant contribution in models with oscillatory potentials in our notation is given by \cite{Chen:2008wn}
\be\label{HI}
H_I(t)\supset -\int d^3x\; a^3(t)\epsilon(t)\dot{\delta}(t)\mathcal{R}^2({\bf x},t)\dot{\mathcal{R}}({\bf x}, t)\,.
\ee
Using equations~\eqref{eq:rx} and~\eqref{eq:rk}, one finds that the contribution to the three-point function from the term~\eqref{HI} in the interaction Hamiltonian is given by
\begin{multline}
\langle\mathcal{R}({\bf k_1},t)\mathcal{R}({\bf k_2},t)\mathcal{R}({\bf k_3},t)\rangle=(2\pi)^3i\,\delta^3({\bf k_1}+{\bf k_2}+{\bf k_3})\mathcal{R}_ {k_1}(t)\mathcal{R}_ {k_2}(t)\mathcal{R}_ {k_3}(t)\times\\\int^t_{-\infty}\;dt'\;2a^3(t')\epsilon(t')\dot\delta(t')\left[\mathcal{R}^*_{k_1}(t')\mathcal{R}^*_{k_2}(t')\dot{\mathcal{R}}^*_{k_3}(t')+2\; perm.\right]+c.c.\,
\end{multline}
Observational constraints on the two-point function imply that knowing the result to linear order in $b_*$ will be enough. The oscillatory nature of $\delta$ is what makes this contribution the dominant one. So to linear order in $b_*$, we can replace $\epsilon$ by $\epsilon_0$, $\delta$ by $\delta_1$ and use the unperturbed mode functions for $\mathcal{R}_ {k}(t)$, {\it i.e.} equation~\eqref{eq:solunpert}.
One might be concerned that the derivative of the correction to the mode function becomes large during the resonance and should be kept, but this is not the case. To see this, notice that there is no contribution to the integral after the modes have frozen out. In fact, we will see that the main contribution arises when the modes are still deep inside the horizon. In this limit, {\it i.e.} for large $x$, the ratio of the absolute value of the time derivative of the unperturbed part $\dot{\mathcal{R}}_{k\,,0}(t)$ and the absolute value of the time derivative of the correction $\dot{\mathcal{R}}_{k\,,1}(t)$ becomes
\be
\frac{|\dot{\mathcal{R}}_{k\,,1}|}{|\dot{\mathcal{R}}_{k\,,0}|}=\left|\cm(x)+i\frac{d}{dx}\cm(x)\right|\,.
\ee  
The results in the last section imply that this is small. Equation~\eqref{exact} tells us that the absolute value of the first term is never significantly larger than $3b_*\sqrt{\pi/2}(f/\sqrt{2\epsilon_*})^{1/2}$, and equation~\eqref{eq:cmprime} reveals that the absolute value of the second term is always less than $6bf/\sqrt{2\epsilon_*}$. As we will see, large non-Gaussianities can only be generated for decay constants satisfying $f\ll\sqrt{2\epsilon_*}$, so that both are small. We conclude that 
\be
\frac{|\dot{\mathcal{R}}_{k\,,1}|}{|\dot{\mathcal{R}}_{k\,,0}|}\ll1\,.
\ee 
In other words, we can use the unperturbed mode functions, and the approximation becomes better for decreasing axion decay constant.
We are interested in the value of the three-point function after horizon exit. In this case, we can replace the factors $\mathcal{R}_ {k_i}(t)$ outside the integral by $\mathcal{R}_ {k_i,0}^{(o)}$, and take the upper limit of the integral to zero. We will drop the dependence of the mode functions on the slow-roll parameters $\epsilon_0$ and $\delta_0$, as well as $\epsilon_1$, and use expression~\eqref{eq:solunpert} for the mode functions inside the integral. This leads to
\be \label{gen}
\langle\mathcal{R}({\bf k_1},t)\mathcal{R}({\bf k_2},t)\mathcal{R}({\bf k_3},   t)\rangle&=&(2\pi)^7\Delta_\mathcal{R}^4\frac{1}{k_1^3k_2^3k_3^3} \delta^3({\bf k_1}+{\bf k_2}+{\bf k_3})\\
&&\hskip-3cm\times\int_0^{\infty}dX \frac{\dot \delta_1}{8H} e^{-i X}
 \left[-i k_1k_2k_3-\frac1X \sum_{{i\neq j}} k_i^2  k_j +\frac{i}{X^2}K(k_1^2+k_2^2+k_3^2) \right] +c.c\,,\nonumber
\ee
where $H$ and $\dot\delta_1$ should be thought of as functions of $X\equiv-K\tau$, and $K\equiv k_1+k_2+k_3$ is the perimeter of the triangle in momentum space. 

When visualizing the results, it is often convenient to introduce a quantity that contains the information about the deviation of the three-point function from scale invariance rather than the three-point function itself. We will use the notation of~\cite{Chen:2008wn} and define\footnote{There is a factor of $9/10$ between our definition of $\mathcal{G}$ and theirs, {\it i.e.} $\mathcal{G}_\text{there}=10\mathcal{G}_\text{here}/9$, which was introduced there presumably to match the WMAP conventions for the local case, where a famous factor of $3/10$ appears. The remaining factor of $3$ arises because the matching is conventionally done in the equilateral limit where three terms become equal.}
\be
\langle\mathcal{R}({\bf k_1},t)\mathcal{R}({\bf k_2},t)\mathcal{R}({\bf k_3},t)\rangle&=&(2\pi)^7\Delta_\mathcal{R}^4\frac{1}{k_1^2k_2^2k_3^2} \delta^3({\bf k_1}+{\bf k_2}+{\bf k_3})\frac{\mathcal{G}(k_1,k_2,k_3)}{k_1k_2k_3}\,.
\ee
It can then be seen from equation~\eqref{gen} that for all models whose three-point function receives its dominant contribution from the term in the interaction Hamiltonian~\eqref{HI}, and for which the mode functions are well approximated by the unperturbed ones, one has
\be\label{eq:shapegen} 
&&\hskip-1cm\frac{\mathcal{G}(k_1,k_2,k_3)}{k_1k_2k_3} =\frac18\int_0^{\infty}dX \frac{\dot \delta_1}{H} e^{-i X}
 \left[-i -\frac1X \sum\limits_{i\neq j} \frac{k_i}{  k_j} +\frac{i}{X^2}\frac{K(k_1^2+k_2^2+k_3^2)}{k_1k_2k_3} \right] +c.c\,.
\ee

If the integral receives its main contribution from a small neighborhood around some value of $X=X_\text{res}$, as is the case in our example, we can replace $1/X$ and $1/X^2$ by $1/X_\text{res}$ and $1/X_\text{res}^2$, respectively, and find that the shape of non-Gaussianity is given by
\be \label{eq:Gres}
&&\hskip-1cm\frac{\mathcal{G}(k_1,k_2,k_3)}{k_1k_2k_3}                          =\frac14
 \left[\text{Im}\,\mathcal{I}_K -\frac{1}{X_\text{res}} \sum\limits_{i\neq j} \frac{k_i}{  k_j}\text{Re}\,\mathcal{I}_K - \frac{1}{X_\text{res}^2}\frac{K(k_1^2+k_2^2+k_3^2)}{k_1k_2k_3}\text{Im}\,\mathcal{I}_K \right]\,,
\ee
where we have defined the integral
\be\label{eq:IKint}
\mathcal{I}_K\equiv\int_0^{\infty}dX \frac{\dot \delta_1}{H} e^{-i X}\,.
\ee

We see that all we need in order to calculate the shape of non-Gaussianities for models of this class is the quantity $\dot\delta_1/H$ as a function of $X$. We have already given this quantity as a function of the scalar field in equation~\eqref{eq:dd1}. It remains to write it as a function of $X$. To leading order in the slow-roll approximation the scalar field is given in terms of $X$ as
\be\label{eq:phiofx}
\phi_0(X)=\phi_K+\sqrt{2\epsilon_*}\ln X\;\;\;\;\text{with}\;\;\;\;\phi_K=\phi_*-\sqrt{2\epsilon_*}\ln K/k_*\,,
\ee
where $\phi_K$ is the value of the scalar field at the time the mode with comoving momentum $K$ exits the horizon.\footnote{See Appendix~\ref{a:gen} for details.} This gives
\be
\frac{\dot \delta_1}{H}=\frac{3b_*\sqrt{2\epsilon_*}}{f}\cos\left(\frac{\phi_K}{f}+\frac{\sqrt{2\epsilon_*}}{f}\ln X\right)\,.
\ee
The three terms in the integral~\eqref{eq:shapegen} can then be recognized as $\Gamma$-functions, and the integral~\eqref{eq:shapegen} can be done analytically. In the regime $f\ll\sqrt{2\epsilon_*}$, in which the resonance occurs deep inside the horizon, we can also use~\eqref{eq:Gres}. 
The integral~\eqref{eq:IKint} then takes the form
\be\label{IKgen}
\mathcal{I}_K=\frac{3b_*\sqrt{2\epsilon_*}}{f}\int\limits_0^\infty dX \;e^{-    iX}\cos\left(\frac{\phi_K}{f}+\frac{\sqrt{2\epsilon_*}}{f}\ln X\right)\,,
\ee
which can be written in terms of $\Gamma$-functions
\begin{equation}\label{eq:IKGamma}
\mathcal{I}_K=\frac{3ib_* \sqrt{2\epsilon_*                                     }}{2f}\left[e^{\frac{\pi\sqrt{2\epsilon_*}}{2f}}\Gamma\left(1+                  i\frac{\sqrt{2\epsilon_*}}{f}\right)e^{i\frac{\phi_K}{f}}+e^{-                  \frac{\sqrt{2\epsilon_*}\pi}{2f}}\Gamma\left(1-i\frac{\sqrt{2\epsilon_*         }}{f}\right)e^{-i\frac{\phi_K}{f}}\right]\,.
\end{equation}
The absolute values of the $\Gamma$-functions in the first and second term are      identical so that the first term dominates for small $f/\sqrt{2\epsilon_*}$ because of the exponential factors. This dominant contribution arises from a neighborhood of size  $(\sqrt{2\epsilon_*}/f)^{1/2}$ around $X_\text{res}=\sqrt{2\epsilon_*}/f$ where the phase of the integrand in equation~\eqref{IKgen} becomes stationary. Equation~\eqref{eq:Gres} can    then be used as long as $f\ll\sqrt{2\epsilon_*}$, which is the regime we are interested   in. Either performing the integral directly in the stationary phase approximation or using Stirling's approximation in equation~\eqref{eq:IKGamma}, one finds that up to a $K$-independent phase
\be
\mathcal{I}_K=-\frac{3 b_*\sqrt{2\pi}}{2}\left(\frac{ \sqrt{2\epsilon_*         }}{f}\right)^{3/2}e^{i\frac{\phi_K}{f}}\,.\ee
Combining this with equation~\eqref{eq:Gres}, we see that the shape of resonant non-Gaussianity is given by
\begin{multline}
\frac{\mathcal{G}(k_1,k_2,k_3)}{k_1k_2k_3}=-\frac{3\sqrt{2\pi}b_*}{8}\left(\frac{\sqrt{2\epsilon_*}}{f}\right)^{3/2}\left[\sin\left(\frac{\phi_K}{f}\right)-\frac{f}{\sqrt{2\epsilon_*}}\sum\limits_{i\neq j} \frac{k_i}{  k_j}\cos\left(\frac{\phi_K}{f}\right)\right.\\\left.-\left(\frac{f}{\sqrt{2\epsilon_*}}\right)^2\frac{K(k_1^2+k_2^2+k_3^2)}{k_1k_2k_3}\sin\left(\frac{\phi_K}{f}\right)\right]\,.
\end{multline}
Other terms in the interaction Hamiltonian also contribute at order $(f/\sqrt{2\epsilon_*})^2$, but these contributions are too small to be phenomenologically interesting and we will drop them. Ignoring a $K$-independent phase, we thus write the final result for the resonant shape as

\be\label{eq:Gresfin}
\frac{\mathcal{G}(k_1,k_2,k_3)}{k_1k_2k_3}&=&f^\text{res}\left[\sin\left(\frac{\sqrt{2\epsilon_*}}{f}\ln K/ k_*\right)\right.\\
&&\quad\quad\quad\left.+\frac{ f}{\sqrt{2\epsilon_*}}\sum\limits_{i\neq j} \frac{k_i}{       k_j}\cos\left(\frac{\sqrt{2\epsilon_*}}{f}\ln K/k_*\right)+\dots\right]\,,\nonumber
\ee
or equivalently for the three-point function
\begin{multline}
\langle\mathcal{R}({\bf k_1},t)\mathcal{R}({\bf k_2},t)\mathcal{R}({\bf k_3},   t)\rangle=(2\pi)^7\Delta_\mathcal{R}^4\frac{1}{k_1^2k_2^2k_3^2} \delta^3({\bf   k_1}+{\bf k_2}+{\bf k_3})\\\times                                               f^\text{res}\left[\sin\left(\frac{\sqrt{2\epsilon_*}}{f}\ln K/k_*\right)+       \frac{f}{\sqrt{2\epsilon_*}} \sum\limits_{i,j} \frac{k_i}{                      k_j}\cos\left(\frac{\sqrt{2\epsilon_*}}{f}\ln K/k_*\right)+\dots\right]\,,
\end{multline}
with
\begin{equation}
f^\text{res}=\frac{3 b_*\sqrt{2\pi}}{8}\left(\frac{\sqrt{2\epsilon_*            }}{f}\right)^{3/2}\,.\label{fresgen}
\end{equation}
The dots stand for terms that have been dropped because they are higher order   in slow-roll or $f/\sqrt{2\epsilon_*}$.

A few comments on this result are in order. Notice that both the frequency of the oscillation and, when written in terms of the monotonicity parameter $b_*$, the amplitude $f^\text{res}$ depend only on $\sqrt{2\epsilon_*}/f$. If this type of non-Gaussianity were measured, it would thus not be possible to distinguish between different potentials from this measurement alone. However, a measurement of the amplitude of tensor modes would give us a direct measurement of $\epsilon_*$ and hence break the degeneracy.\footnote{For a general discussion of the implications of a measurement of tensor modes for the inflationary theory see~\cite{Baumann:2008aq} and references therein.}

Concerning detectability, notice also that when the axion decay constant becomes too small, the frequency of the oscillations becomes too high to be experimentally resolvable. Let us assume the signal can be resolved if the period is longer than $\Delta \ell\sim1$ around the first Doppler peak, {\it i.e.} near $\ell\sim200$.\footnote{This will then also be true for all larger $\ell$}. The periodicity near a given value of $\ell$ is given by $2\pi \ell f/\sqrt{2\epsilon_*}$. So our condition is $\sqrt{2\epsilon_*}/f\lesssim 2\pi \ell $, and taking $\ell=200$ leads to  
\be\label{44b}
f^\text{res}\lesssim 4 \times 10^4\, b_*\,.
\ee

In order to go from this to a numerical value for the upper bound on $f^\text{res}$, one needs an upper bound on $b_*$ from comparison of the predicted power spectrum with the data. So far this comparison has only been done for the linear potential \cite{Flauger:2009ab}, which will be the subject of the next subsection, and the quadratic potential \cite{Pahud:2008ae}.\footnote{For another observational constraint on an oscillatory power spectrum due to a deviation from the Bunch-Davies vacuum see~\cite{Okamoto:2003wk}.} However, in \cite{Pahud:2008ae}, only decay constants $f$ ($\beta \Mpl$ in their notation) larger than $5\times 10^{-3}$ were considered, which is too large to give a sizable $f^{\text{res}}$.
  
It may also be interesting to consider a potential of the form
\be
V(\phi)=V_0(\phi)\left[1+\lambda\cos\left(\frac{\phi}{f}\right)\right]\,.
\ee
In the approximation we have been working in, our results can immediately be translated to this type of potential by replacing $b_*f\sqrt{2\epsilon_*}\to\lambda$. 
One finds that\footnote{See Appendix~\ref{a:gen} for details.}
\be
f^\text{res}=\frac{3\lambda\sqrt{2\pi}}{8f^2} \left(\frac{\sqrt{2\epsilon_*}}{f}\right)^{1/2}\,.
\ee
This type of potential with $V_0(\phi)=\frac12m^2\phi^2$ was studied by Chen, Easther, and Lim in~\cite{Chen:2008wn}. One has
\be
\sqrt{2\epsilon_*}=\frac{2}{\phi_*}\,,
\ee
so that our result is
\be
f^\text{res}=\frac{3\sqrt{\pi}}{4}\frac{\lambda}{f^{2}\sqrt{f\phi_*}}\,.
\ee
This is smaller than their analytic estimate for the amplitude in the           equilateral limit by a factor of $10\sqrt{\pi}/27$, which is in agreement with their statement that their analytic estimate overpredicts the numerical result by about $30\%$.\footnote{For the comparison, recall that there is difference by a factor of $10/9$ between their definition of $f_\text{res}$ and our $f^\text{res}$ from $\mathcal{G}_\text{there}=10\mathcal{G}_\text{here}/9$.}

%%%%%%%%%%%%%%%%%%%%%%%%%%%%%%%%%%%%%%%%%%%%%%%%%%%%%%%%%%%%%%%%%%%%%%%%%%%%%%%%%%%%%%%%%%%%%%%%%%%%%%%%%%%%%

\subsection{The bispectrum for a linear potential}
We will now consider axion monodromy inflation~\cite{McAllister:2008hb, Flauger:2009ab} in some detail as a special case.
The low energy effective theory describing the system is that of a canonically normalized real scalar field with potential
\be\label{eq:V} 
V(\phi)=\mu^3\phi+\Lambda^4\cos \left(\frac{\phi}{f}\right)=\mu^3\left[\phi+b   f\cos \left(\frac{\phi}{f}\right)\right]\,.
\ee
The parameter $\mu$ has dimensions of a mass and is fixed by COBE normalization to be approximately $\mu\simeq 6\times 10^{-4}$.

For this potential, one finds $\sqrt{2\epsilon_*}=1/\phi_*$, and $b_*=b$ is now independent of $\phi_*$. We thus conclude that the shape of non-Gaussianities in this model is 
\begin{equation}\label{eq:Gresfinlin}
\frac{\mathcal{G}(k_1,k_2,k_3)}{k_1k_2k_3}=f^\text{res}\left[\sin\left(\frac{\ln K/k_*}{f\phi_*}\right)+f\phi_*\sum\limits_{i\neq j} \frac{k_i}{  k_j}\cos\left(\frac{\ln K/k_*}{f\phi_*}\right)+\mathcal{O}\left((f\phi_*)^2\right)\right]\,,
\end{equation}
or equivalently for the three-point function 
\begin{multline}
\langle\mathcal{R}({\bf k_1},t)\mathcal{R}({\bf k_2},t)\mathcal{R}({\bf k_3},t)\rangle=(2\pi)^7\Delta_\mathcal{R}^4\frac{1}{k_1^2k_2^2k_3^2} \delta^3({\bf k_1}+{\bf k_2}+{\bf k_3})\\\times f^\text{res}\left[\sin\left(\frac{\ln K/k_*}{f\phi_*}\right)+f\phi_* \sum\limits_{i,j} \frac{k_i}{                k_j}\cos\left(\frac{\ln K/k_*}{f\phi_*}\right)+\mathcal{O}\left((f\phi_*)^2\right)\right]\,,
\end{multline}
where 
\begin{equation}\label{eq:fres}
f^\text{res}=\frac{3\sqrt{2\pi}b}{8(f\phi_*)^{3/2}}\,.
\end{equation}
Once again, this result is valid at leading order in an expansion in $b$, which is constrained to be much less than one by comparison of the two point function with CMB data \cite{Flauger:2009ab}.\footnote{This is true provided the potential is monotonic, {\it i.e.} $b<1$.} $\phi_K$ is the value of the field at which modes with comoving momentum $K=k_1+k_2+k_3$ exit the horizon, $\phi_*$ is the value of the inflaton field when the pivot scale $k_*$ exits the horizon. Provided this happens around 60 $e$-folds before the end of inflation, one has $\phi_*\sim 11$. Finally the axion decay constant $f$ is a free parameter. A typical value in an explicit string construction is $10^{-4}\lesssim f\lesssim 10^{-1}$. The result is valid provided $f \phi_*\ll 1$. Logarithmic dependences on $k$ which arise from the dependence of the mode functions on $\epsilon_0$ as well as $\delta_0$ were neglected. 

As the axion decay constant decreases, the frequency of the oscillations increases linearly and, keeping the parameter $\Lambda$ in the potential~\eqref{eq:V} fixed, the amplitude increases rapidly  like $f^{5/2}$. It is a natural question to ask for what values of the axion decay constant an observably large signal can be generated in this model while satisfying the bounds on the power spectrum from the data. As long as $f<10^{-2}$, the bound at 95\% confidence level is summarized approximately by $bf<10^{-4}$~\cite{Flauger:2009ab}.
Combining this with equation~\eqref{eq:fres}, one finds that
\be
f^\text{res}\lesssim\frac{3\sqrt{2\pi}}{8f^{5/2}\phi_*^{3/2}}\times 10^{-4}\simeq\left(\frac{6\times 10^{-3}}{f}\right)^{5/2}\,.
\ee
The regime in which the model can simultaneously be consistent with the constraints on the two-point function and generate an observably large three-point function is thus $f\lesssim 6\times 10^{-3}$.
As we mentioned for the general case, requiring that the period of the oscillation should be larger than $\Delta\ell\sim 1$ for $\ell\approx200$ leads to a lower bound on $f$. For the linear case, it is $f\gtrsim 10^{-4}$. The range of axion decay constants for which observably large non-Gaussianities can be generated is thus approximately
\be\label{frange}
10^{-4}\lesssim f\lesssim 6\times 10^{-3}\,.
\ee
Notice that in this range $f\phi_*\ll1$ is always satisfied.

\begin{figure}[h]
\begin{center}
\includegraphics[width=5in]{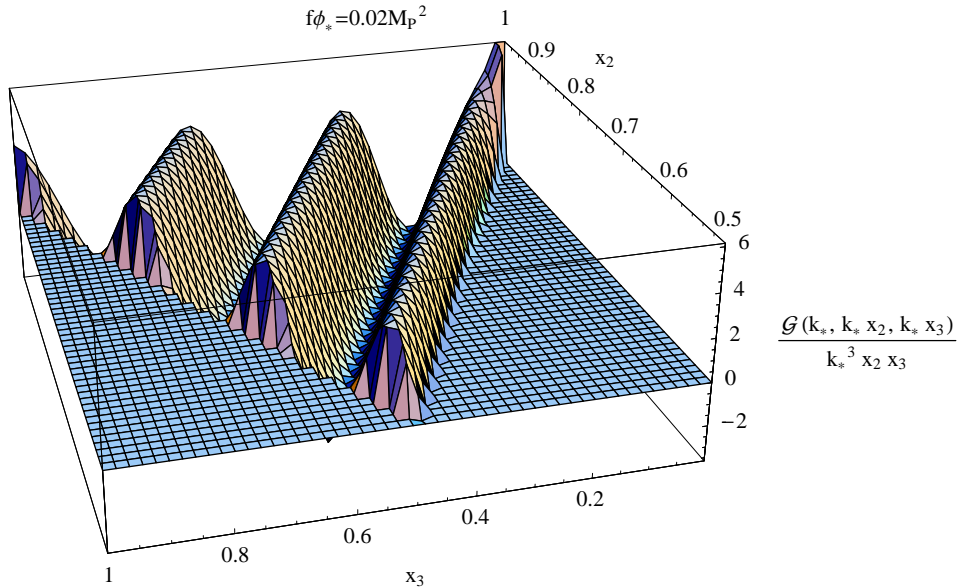}
        \caption{This plot shows the shape $\mathcal{G}(k_1,k_2,k_3)/(k_1k_2k_3)$ of resonant non-Gaussianity for the linear potential of axion monodromy inflation with $b=10^{-2}$, $f\phi_*=2\times10^{-2}$ and fixed $k_1=k_*=0.002\,\text{Mpc}^{-1}$. We use the notation $x_2=k_2/k_1$ and $x_3=k_3/k_1$. The triangle inequality implies $x_2+x_3\leq 1$ and the quantity is symmetric under interchange of $x_2$ and $x_3$ so that we show in the plot only the region $1/2 \leq x_2 \leq 1$.}
      \label{fig:shape}
\end{center}
\end{figure}

The shape of resonant non-Gaussianity for axion monodromy inflation is shown in Figure~\ref{fig:shape} for $b=10^{-2}$, $f\phi_*=2\times10^{-2}$, and fixed $k_1=k_*=0.002\,\text{Mpc}^{-1}$. We chose this value of $f$ because both the leading contribution and the subleading contribution in $f\phi_*$ are clearly visible. Notice that as the value of $k_1$ changes, the phase of the oscillation changes. 

We find that our analytic result for $f^\text{res}$ agrees with the values obtained by numerical integration in~\cite{Hannestad:2009yx} at the per cent level.\footnote{For the comparison, notice that~\cite{Hannestad:2009yx} uses a momentum dependent quantity $\tilde{f}_{NL}$. In the equilateral limit, they extract their quantity $f_A=-\tilde{f}^{(eq)}_{NL}$. This quantity is related to our $f^\text{res}$ according to $f_A=10f^\text{res}/9$.}

%%%%%%%%%%%%%%%%%%%%%%%%%%%%%%%%%%%%%%%%%%%%%%%%%%%%%%%%%%%%%%%%%%%%%%%%%%%%%%%%%%%%%%%%%%%%%%%%%%%%%%%%%%%%%

\subsection{Consistency relation}

As pointed out in \cite{Maldacena:2002vr} (see also \cite{Creminelli:2004yq}), in the limit in which one of the momenta, say, $k_3$ is much less than the other two, which are then roughly equal, $k_3\ll k_1\approx k_2=k$, the three-point function is related to the two-point function by a consistency relation
\be \label{cons}
\lim_{k_3\rightarrow 0}\langle\mathcal{R}({\bf k_1},t)\mathcal{R}({\bf k_2},t)\mathcal{R}({\bf k_3},t)\rangle\simeq-|\mathcal{R}_{k_3}^{(o)}|^2\frac{1}{H(t_k)}\frac{d}{dt_k}\langle\mathcal{R}({\bf k_1},t)\mathcal{R}({\bf k_2},t)\rangle \,,
\ee
where $t_k$ is the time at which $k_1\approx k_2= k$ exit the horizon, and we will take $t$ to be some late time when the modes with comoving momenta $k_1$, $k_2$ and $k_3$ have exited the horizon. In our conventions the two-point function is given by
\be
\langle\mathcal{R}({\bf k_1},t)\mathcal{R}({\bf k},t)\rangle=(2\pi)^3\delta({\bf k_1}+{\bf k})|\mathcal{R}_{k}^{(o)}|^2\,.
\ee
Recall that 
\be
\left|\mathcal{R}_k^{(o)}\right|^2=2\pi^2 \frac{\Delta_\mathcal{R}^2(k)}{k^3}\,,
\ee
and remember that $\Delta_\mathcal{R}^2(k)$ is momentum independent in the scale-invariant limit. It acquires its momentum dependence from the time dependence of the background. Modes with different momenta feel a different background as they exit the horizon. In equation~\eqref{cons}, we can thus make the replacement
\be
\frac{1}{H(t_k)}\frac{d}{dt_k}\to\frac{d}{d\ln k}\,,
\ee
with the derivative in $\ln k$ only acting on the momentum dependence in $\Delta_\mathcal{R}^2(k)$. Restoring the ${\bf k_3}$ dependence inside the $\delta$-function, the consistency relation can then be written in the form
\be\label{eq:conscond}
\lim_{k_3\rightarrow 0}\langle\mathcal{R}({\bf k_1},t)\mathcal{R}({\bf k_2},t)\mathcal{R}({\bf k_3},t)\rangle\simeq-(2\pi)^3\delta^3({\bf k_1}+{\bf   k_2}+{\bf k_3})|\mathcal{R}_{k_3}^{(o)}|^2|\mathcal{R}_{k}^{(o)}|^2 \frac{d\ln\Delta_\mathcal{R}^2(k) }{d\ln k}\,.
\ee
The logarithmic derivative of the amplitude of scalar fluctuations~\eqref{eq:d2r} is given by
\begin{equation}
\frac{d \ln \Delta_\mathcal{R}^2}{d \ln k}\simeq n_s-1+\delta n_s\frac{\sqrt{2\epsilon_*}}{f} \sin \left(\frac{\phi_k}{f}\right)\;\;\;\;\text{with}\;\;\;\;\delta n_s=3b_*\left( \frac{2\pi f}{\sqrt{2\epsilon_*}}\right)^{1/2}\,.
\end{equation}
The consistency condition~\eqref{eq:conscond} together with the power spectrum given in Section 2 implies that the shape in this limit up to a phase in the trigonometric function should take the form
\begin{equation}
\frac{\mathcal{G}(k,k,k_3)}{k^2k_3}=\frac{3\sqrt{2\pi}b_*}{8}\left(\frac{f}{\sqrt{2\epsilon_*}}\right)^{1/2}\frac{2k}{k_3}\cos\left(\frac{\sqrt{2\epsilon_*}}{f}\ln 2k/k_*\right)\,.
\end{equation}
This agrees with our result for the resonant shape~\eqref{eq:Gresfin} after setting $k_1=k_2=k$ and taking the limit $k_3\ll k$.

Notice that in~\cite{Hannestad:2009yx}, the consistency relation \eqref{cons} was used to predict the shape of resonant non-Gaussianities in the squeezed limit. Here we have derived the resonant shape from first principles, and we use the consistency relation as a check of our computation.

%%%%%%%%%%%%%%%%%%%%%%%%%%%%%%%%%%%%%%%%%%%%%%%%%%%%%%%%%%%%%%%%%%%%%%%%%%%%%%%%%%%%%%%%%%%%%%%%%%%%%%%%%%%%%

\section{Correlations Between Resonant and Other Types of Non-Gaussianities}\label{s:cosine}

The comparison of theoretical models of non-Gaussianity with the data is computationally very challenging. In light of this difficulty, the authors of \cite{Babich:2004gb} have proposed to use a normalized scalar product, or ``cosine'', to assess to which extent two different 3D primordial shapes give rise to a similar 2D signal in the CMB. If two shapes have a large cosine (in absolute value), the observational constraints on one can be exported to the other.

Following this idea, in \cite{Fergusson:2008ra}, an appropriately defined cosine has been used to classify known non-Gaussian models. One class was reserved for models of canonically normalized single-field inflation where some feature is present on the top of an otherwise slow-roll flat potential. The models considered in this work belong to this class.

In this section we show that the shape of resonant non-Gaussianity that we have derived is very different from the shapes of non-Gaussianity that have been constrained by data. In particular we compute the correlation (to be defined soon) between resonant non-Gaussianity and local, equilateral and orthogonal non-Gaussianity and find that it is always less than about $10\%$. The observational constraints on these models are therefore not useful to constrain resonant non-Gaussianity.

%%%%%%%%%%%%%%%%%%%%%%%%%%%%%%%%%%%%%%%%%%%%%%%%%%%%%%%%%%%%%%%%%%%%%%%%%%%%%%%%%%%%%%%%%%%%%%%%%%%%%%%%%%%%%%

\subsection{Scalar product, cosine and shapes}

In this subsection we give the definition of the cosine that we will use in the next subsections to compute the correlation between resonant non-Gaussianity and local, equilateral and orthogonal shapes.   

Following \cite{Fergusson:2008ra}, we choose the simplest product that exhibits the same scaling as the optimal CMB estimator. The definition is
\be \label{sp}
F(S,S')=\int_{\V} S(k_1,k_2,k_3) S'(k_1,k_2,k_3) \frac{d\V}{K}\,,
\ee
where $d\V=dk_1dk_2dk_3$ and
\be
S(k_1,k_2,k_3)\propto  \frac{\mathcal{G}(k_1,k_2,k_3)}{k_1k_2k_3}\,.
\ee
The normalization is irrelevant for our purpose because we will only be interested in the normalized scalar product or cosine of two shapes given by
\be\label{cosine}
C(S,S')\equiv\frac{F(S,S')}{\sqrt{F(S,S)F(S',S')}}\,.
\ee
Because of the rotational and translational symmetries of the background geometry, the volume of integration $\V$ is three dimensional. The volume three-form $d\V$ and the integration boundaries are conveniently written in the coordinates $\left\lbrace k,\alpha,\beta \right\rbrace$, which are related to $\left\lbrace k_1,k_2,k_3\right\rbrace$ by \cite{Fergusson:2006pr}
\be
k\equiv\frac{K}{2}=\frac12\left(k_1+k_2+k_3\right) \,, \quad k_1\equiv k\left(1-\beta\right)\,,\\
k_2\equiv \frac k2\left(1+\alpha+\beta\right)\,,\quad k_3\equiv \frac k2 \left(1-\alpha+\beta\right)\,.
\ee
One virtue of this set of coordinates is that for scale-invariant non-Gaussian models, {\it i.e.} $\mathcal{G}(k,k,k)\propto k^3$ or $S(k,k,k)\propto k^{0}$, the integrals over $dk$ cancel between the numerators and denominators in \eqref{cosine} and we are left with a two dimensional integral. However, resonant non-Gaussianity is not scale invariant (due to the sine and cosine in {\it e.g.} \eqref{Sres}) and having a three-dimensional integration volume is essential to get meaningful results for the cosine \eqref{cosine}.

The integration boundaries in \eqref{sp} require some discussion. In \cite{Fergusson:2008ra}, they were chosen to be $0\leq k\leq \infty$, $-(1-\beta)\leq\alpha\leq(1-\beta)$ and $0\leq \beta\leq1$. Notice that $\alpha$ and $\beta$ parametrize a triangle that is the base of a tetrahedron in the space $\left\lbrace k_1,k_2,k_3\right\rbrace$ with the apex at the origin and the semi-perimeter $k$ parameterizing the height. With the above integration limits, the product \eqref{sp} is typically infinite. Depending on the shapes that are being integrated, there can be an IR divergence for example where $\alpha=\pm 1$ or $\beta=1$, {\it i.e.} at the vertices of the $\left\lbrace \alpha,\beta \right\rbrace$ triangle where one of the $k_i$ vanishes. Local non-Gaussianity \eqref{loc} and resonant non-Gaussianity \eqref{Sres} show this divergence for squeezed configurations. In addition there are generically also UV divergences from the $dk$ integral\footnote{As we said, the UV divergence can be neglected in the case of scale-invariant shapes because the $dk$ integral simplifies in the cosine \eqref{cosine}.}.

Physically it is clear that for a given experiment, {\it e.g.}~observations of the CMB or of large scale structure (LSS), there is a finite range of momenta $\left\lbrace \kin,\kax\right\rbrace$ that can be probed. The cosine \eqref{cosine} is useful if it compares two primordial non-Gaussian shapes over the same range of momenta that is probed by a chosen class of experiments. Implementing this is slightly subtle because of the three-dimensional nature of the primordial non-Gaussian shapes as opposed to the two-dimensional nature of the observations (see {\it e.g.} \cite{Babich:2004gb,Fergusson:2008ra} for a discussion). The interesting issue of defining a scalar product suitable for non-scale-invariant shapes is beyond the scope of this work, so we will limit ourselves to specify some $\left\lbrace \kin,\kax\right\rbrace$ range of integration for $k_i$ and check that our results do not qualitatively depend on this choice.

Let us now turn to the shapes that we will consider. For resonant non-Gaussianity, we will work with the linear potential derived from the string theoretic construction. It is clear from \eqref{cosine} that the normalization of $S$ is irrelevant. We can thus define
\be\label{Sres}
S_\text{res}(k_1,k_2,k_3)\equiv\sin\left(\frac{\ln K}{f\phi_*}\right)+f\phi_*\cos\left(\frac{\ln K}{f\phi_*}\right) \sum_{i\neq j}\frac{k_i}{ k_j}\,.
\ee
We would like to know the correlation of $S_\text{res}$ with other shapes that have already been compared to and constrained by observations. If the cosine were close to one for some of them, we could export their constraints to resonant non-Gaussianity. The best constraints form 7-year WMAP data on local, equilateral, and orthogonal non-Gaussianity at $95\%$ CL are~\cite{Komatsu:2010fb}\footnote{For optimal limits from 5-year WMAP data see~\cite{Smith:2009jr,Senatore:2009gt}.}
\be
-10<f^\text{local}<74\,,\quad -214<f^\text{equil}<266\,,\quad -410<f^\text{ortho}<6\,.
\ee
These shapes are defined by
\be
S_\text{local}(k_1,k_2,k_3)&\equiv&\frac{k_1^3+k_2^3+k_3^3}{k_1 k_2 k_3}\,,\label{loc}\nonumber\\
S_\text{equil}(k_1,k_2,k_3)&\equiv&\frac{(k_1+k_2-k_3)(k_1+k_3-k_2)(k_3+k_2-k_1)}{k_1k_2k_3}\\\nonumber
&=&\left[-\frac{ k_3^2}{k_1k_2}-\frac{ k_1^2}{k_3 k_2}-\frac{ k_2^2}{k_1k_3}-2+\sum_{i\neq j}\frac{k_i}{ k_j}\right]\,,\\
S_\text{ortho}(k_1,k_2,k_3)& \equiv & 3 S_\text{equil}(k_1,k_2,k_3)-2\,,\non
&=&\left[-\frac{3 k_3^2}{k_1k_2}-\frac{3 k_1^2}{k_3 k_2}-\frac{3 k_2^2}{k_1k_3}-8+3\sum_{i\neq j}\frac{k_i}{ k_j}\right]\,.\nonumber
\ee
Notice that these shapes are factorizable approximations to the results of the theoretical calculations (the reader is referred to \cite{Senatore:2009gt} for further details). They are good approximations in the sense that their cosine with the theoretical shapes is very close to one.

%%%%%%%%%%%%%%%%%%%%%%%%%%%%%%%%%%%%%%%%%%%%%%%%%%%%%%%%%%%%%%%%%%%%%%%%%%%%%%%%%%%%%%%%%%%%%%%%%%%%%%%%%%%

\subsection{Numerical results}

\begin{figure}[ht]
\begin{center}
\includegraphics[width=6.6in]{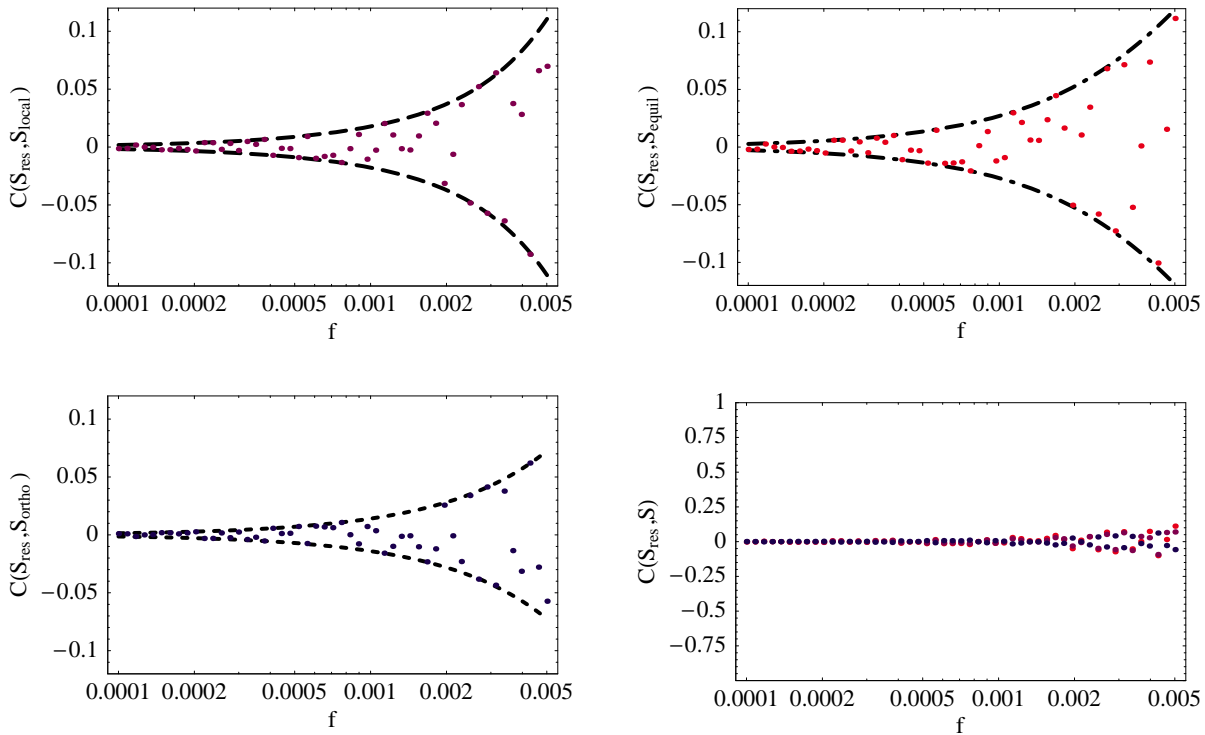}
        \caption{From the bottom left panel clockwise we plot the cosine between the resonant shape \eqref{Sres}, and orthogonal, local and equilateral shapes \eqref{loc}, evaluated numerically for 50 points equally spaced in $\log f$. We also plot the enveloping profiles from the semi-analytical calculation to show that numerical and semi-analytical approaches give consistent results. In the bottom right panel, numerical results for all shapes are shown together. This makes it graphically clear that resonant non-Gaussianity is essentially orthogonal to the other shapes.\label{fig:num}}
\end{center}
\end{figure}

We have numerically calculated the cosine \eqref{cosine} between the shape of resonant non-Gaussianity \eqref{Sres} and local, equilateral and orthogonal shapes in \eqref{loc}. We have chosen $\kin=10^{-4} \,\text{Mpc}^{-1}$ and $\kax=10^{-1}\,\text{Mpc}^{-1}$ for the IR and UV cutoff, respectively. In order to implement the IR cutoff in the numerical calculation we have introduced three exponential damping factors $\exp(-\kin/k_i)$ with $i=1,2,3$ in the integral \eqref{sp}. The UV cutoff is taken into account using the integration region $0<k\leq 3\kax/2$. The results of this brute force approach are shown in Figure \ref{fig:num}. We have plotted the value of the three cosines as function of the axion decay constant $f$. 

It is clear from the figure that resonant non-Gaussianity has a correlation smaller than about $10\%$ with the other shapes for any interesting value\footnote{The interesting $f$ range was discussed around \eqref{frange}.} of $f$. The cosines take both positive and negative values and get closer to zero as $f$ decreases. Both these features can be understood analytically and this is the subject of the next subsection.

%%%%%%%%%%%%%%%%%%%%%%%%%%%%%%%%%%%%%%%%%%%%%%%%%%%%%%%%%%%%%%%%%%%%%%%%%%%%%%%%%%%%%%%%%%%%%%%%%%%%%%%%%%%

\subsection{Semi-analytical results}

\begin{figure}[ht]
\begin{center}
\includegraphics[width=6.6in]{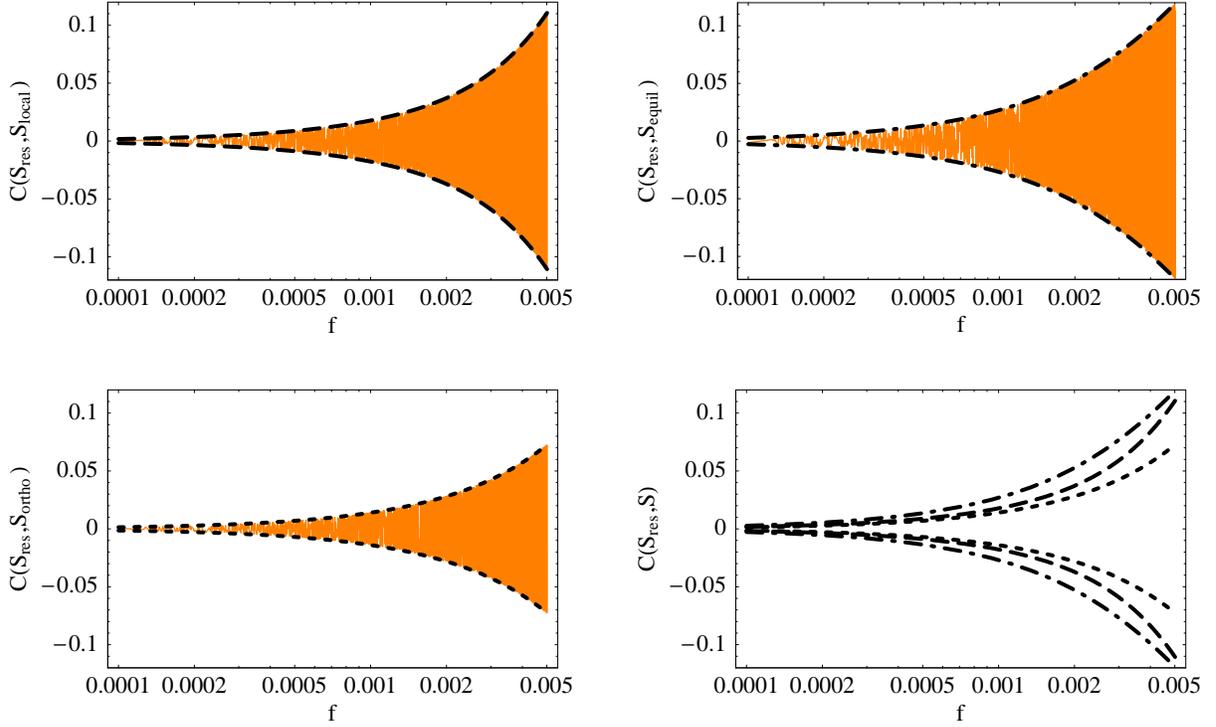}
        \caption{From the bottom left panel clockwise we plot the cosine between the resonant shape \eqref{Sres}, and orthogonal, local, and equilateral shapes \eqref{loc}, as a function of $f$. In the bottom right panel, showing all the three embedding profiles at the same time, makes it graphically clear that resonant non-Gaussianity has less than about $10\%$ correlation with the other shapes. A linear fit to the enveloping curves is given in \eqref{numfit}.\label{fig:ana}}
\end{center}
\end{figure}

To better understand the points in Figure \ref{fig:num}, we have calculated the three cosines semi-analytically. We adopted a simplification in implementing the IR cutoff $k_i>\kin$: we took as integration region $-(1-\beta)<\alpha<1-\beta$ and $\kin/\kax\leq \beta\leq 1-\kin/\kax$. This cuts off not only the vertices of the $\left\lbrace\alpha,\beta\right\rbrace$ triangle but also the side of the triangle at $\beta=0$. Given that none of the shapes we are considering has a divergent contribution along a side (as would be {\it e.g.}~the case for flat shapes obtained in the presence of deviations from a Bunch-Davies vacuum \cite{Holman:2007na,Meerburg:2009fi}), the result we get is a good approximation to the one obtained from cutting off only the vertices. As can be seen in Figure \ref{fig:num}, the results of this subsection nicely agree with those of the numerical approach described above, in which only the vertices had been cut off by the exponential damping factors $\exp(-\kin/k_i)$. We have also checked that varying the cutoff in $\beta$ away from $10^{-3}$ by up to an order of magnitude changes the value of the cosines by less than a per cent. The cosines plotted in Figure \ref{fig:ana} as function of $f$ are computed using the above integration boundaries for $\alpha$ and $\beta$ and an integration range $3\kin \leq k \leq 3 \kax /2$ for $k$.

The very rapid oscillations are due to the $k$-independent phase $\phi_*/f$ and do not have a particular physical relevance given that the phase of the oscillations is arbitrary in the present model. On the other hand, the enveloping profiles, which are highlighted in Figures~\ref{fig:num} and \ref{fig:ana}, carry some interesting information. They show that resonant non-Gaussianity has a correlation smaller than about $10\%$ with local, equilateral and orthogonal models. The observational constraints on the latter are hence of little use in constraining resonant non-Gaussianity, the more so the smaller $f$. The enveloping profiles are well described by a linear fit that gives
\be \label{numfit}
|C(S_{res},S_\text{local})|&\leq& 1.9 f \phi_*\,,\\
|C(S_{res},S_\text{equil})|&\leq& 2.3 f \phi_*\,\\
|C(S_{res},S_\text{ortho})|&\leq& 1.3 f \phi_*\,.
\ee
This dependence can be understood as follows. Let us consider the cosine $C(S_{res},S)$ for some slowly varying shape $S$. There are three scalar products appearing in the definition \eqref{cosine} of $C(S_{res},S)$. Because of the oscillations of $S_{res}$, the scalar product $F(S_{res},S)$ can get a contribution from at most  half a period, because an integration over one or more whole periods gives approximately zero. Notice that for any scalar product the largest contribution comes from the region around $k\sim \kax$ because the overall scaling of \eqref{sp} is $F(S,S')\sim \kax^2$. The periodicity of $S_{res}$ as function of $k$ around $k\sim\kax$ is given by $2\pi f \phi_* \kax$. Hence $F(S_{res},S)$ is obtained effectively from integrating over a range of $k$ that is smaller than half a period, \textit{i.e.}~$\pi f \phi_* \kax$. On the other hand, the scalar products $F(S_{res},S_{res})$ and $F(S,S)$ have either squared oscillations or no oscillations at all, respectively. This means that there is now no cancellation due to the oscillations and $F(S_{res},S_{res})$ and $F(S,S)$ get integrated over the whole range of $k$, {\it i.e.} approximately $\kax$. Taking the ratio as in \eqref{cosine}, we see that the absolute value of the cosine $C(S_{res},S)$ can be bounded from above by $\pi f\phi_*$, up to a number smaller than but of order one. Let us say it in other words. If there were no oscillations, $S_{res}$ would be roughly approximated by some linear combination of local and equilateral shapes. Then, always ignoring oscillations, $C(S_{res},S)$ would generically be smaller than but of order one. The presence of oscillations gives the leading effect on the numerator $F(S_{res},S)$ where the range of the $dk$ integral is reduced by a factor $\pi f \phi_*$. Hence, we find again 
\be
|C(S_{res},S)|&\lesssim \pi f \phi_*\,.
\ee
The result of this heuristic argument nicely agrees with the fit of the semi-analytic computation presented in \eqref{numfit}. Notice that the argument given above applies to the cosine of resonant non-Gaussianity with any slowly varying non-Gaussian shape and not just with those considered here.

\section{Discussion}\label{s:disc}

We have studied the primordial bispectrum of scalar perturbations for models whose potential possesses small modulations. 
We do not deny that our work was largely motivated by a class of models derived from string theory that are based on axion monodromy in which such periodic modulations on top of an otherwise flat potential are a generic feature \cite{McAllister:2008hb, Flauger:2009ab,Berg:2009tg}. However, we argue that these are by no means the only models where such oscillations are expected to arise. In large field models of inflation, the inflaton potential is required to be flat over a range in field space much larger than $\Mpl$. From the point of view of effective field theory, a potential that is flat over such a large range seems unnatural unless there is an underlying shift symmetry. This makes axions natural candidates for the inflaton especially in the context of large field inflation. If the inflaton is an axion, it seems plausible that the potential will receive small periodic contributions from non-perturbative effects. These periodic contributions might be due to instantons in a gauge sector the axion couples to, or, in the context of string theory, they might arise from Euclidean branes or world-sheet instantons. String inspired or not, as soon as we use the shift symmetry of axions to explain why the inflaton potential is so flat, we should admit the possibility of small periodic modulations in the potential which may lead to observational consequences.  
So if theoretical prejudices have to be employed to isolate a handful of shapes of non-Gaussianity that should be looked for in the data, resonant non-Gaussianity deserves to be one of them. We hope that, even though it is not factorizable, the analytical expression for the shape of resonant non-Gaussianity given in this work will make it possible to obtain observational bounds on this type of non-Gaussianity. 

The CMB data is compatible with a small logarithmic running of the power spectrum of primordial fluctuations. This has made it natural to consider phenomenological types of non-Gaussianity that deviate from scale invariance by at most a small logarithmic running. On the other hand, the theoretical considerations expressed above suggest that we should keep in mind the possibility of a scale dependence that reproduces scale invariance only after an adequate average.
 
Resonant non-Gaussianity typically comes with oscillations in the two-point function which are constrained by observations \cite{Flauger:2009ab}. Remarkably, keeping $b_*$ fixed, the amplitude of modulations in the two- and three-point function scales in opposite directions when varying the frequency. This implies that if oscillations were really imprinted on cosmological perturbations during inflation, they could equally well become observationally accessible in the two-point function, the three-point function, or in both. This disentanglement of two- and three-point functions is a peculiar feature of resonantly produced perturbations.

In~\cite{Cheung:2007st}, an effective field theory for the fluctuations around a quasi de Sitter background was constructed for the case of a single field. As is well known, for a single scalar field coupled to gravity, one can fix a gauge in which the fluctuations in the scalar field vanish, or are eaten up by the metric. As pointed out in~\cite{Cheung:2007st}, this gauge is very much like unitary gauge in a spontaneously broken gauge theory where the Goldstone has become the longitudinal mode of the gauge field. Also very much like in the gauge theory example, the longitudinal mode, or the Goldstone boson dominates the dynamics at high energies so that the non-linear theory of the Goldstone contains all the information about the system at sufficiently high energies. The theory of the Goldstone, even though non-linear, is easier to study and in particular makes relations between different operators transparent that would otherwise be obscure. The effective field theory of inflation as presented in~\cite{Cheung:2007st} or~\cite{Weinberg:2008hq} is the tool of choice if one is interested in high energy corrections, {\it i.e.} higher derivative corrections. 
However, this high energy limit is equivalent to a limit in which all slow-roll parameters are taken to zero. Since the effect of resonant non-Gaussianity arises from a term in the interaction Hamiltonian that is higher order in the slow-roll expansion, it should not be surprising that it is not captured by the simplest version of the effective field theory of inflation. These terms could of course be kept~\cite{Cheung:2007sv}, but essentially at the cost of turning the effective field theory of inflation back into the system of a single scalar field coupled to gravity that we have studied here, written in a slightly different notation.

Finally, let us conclude with a couple of interesting directions for future research. We have focused our efforts on the three-point function in this work because it is the obvious observable to look for when looking for a departure from Gaussianity. The four-point function may also be of phenomenological interest in these models, and it can be calculated by the same methods presented here. 

Our calculations have shown that the three-point function of primordial curvature perturbations may be large in models with periodically modulated potentials. For a comparison with the data, it still remains to calculate the prediction of the model for the two-dimensional image of the cosmic microwave background. We have seen that our shape is not factorizable. This makes a direct numerical evaluation too time consuming. However, it may be possible to make analytic progress in this direction. 

Constraints on resonant non-Gaussianity could also arise from its effect on large scale structures. A very preliminary analysis shows that the effect on the halo bias discussed in \cite{Verde:2009hy} is relatively modest because it comes from the signal in squeezed configuration which is suppressed in our model by the small factor $f/\sqrt{2\epsilon_*}$. It would be interesting to consider other large scale structure observables.

%%%%%%%%%%%%%%%%%%%%%%%%%%%%%%%%%%%%%%%%%%%%%%%%%%%%%%%%%%%%%%%%%%%%%%%%%%%%%%%%%%%%%%%%%%%%%%%%%%%%%%%%%%%%%

\section*{Acknowledgments}

It is a pleasure to thank Richard Easther, Eiichiro Komatsu, Michele Liguori, Eugene Lim, Liam McAllister, Emiliano Sefusatti, and Gang Xu for many useful comments and discussions. The work of R.F. has been partially supported by the National Science Foundation under Grant No. NSF-PHY-0747868 and the Department of Energy under Grant No. DE-FG02-92ER-40704. The research of E.P. was supported in part by the
National Science Foundation through grant NSF-PHY-0757868.

%%%%%%%%%%%%%%%%%%%%%%%%%%%%%%%%%%%%%%%%%%%%%%%%%%%%%%%%%%%%%%%%%%%%%%%%%%%%%%%%%%%%%%%%%%%%%%%%%%%%%%%%%%%%%

\appendix

\section{Background Solution for the General Potential}\label{a:gen}

In this appendix, we derive the background solution and the slow-roll parameters for a potential of the form
\be\label{eq:Vadd} 
V(\phi)=V_0(\phi) +\Lambda^4 \cos \left(\frac{\phi}{f}\right)\,,
\ee
where we assume $V_0(\phi)$ to be a smooth featureless potential that admits slow-roll inflation. For this potential, the equation of motion for the inflaton becomes
\be
\ddot\phi+3H\dot\phi+V_0'(\phi)=\frac{\Lambda^4}{f}\sin\left(\frac{\phi}{f}\right)\,.
\ee
Since oscillations have not yet been observed, we know that the modulations must at least be small around the values the inflaton takes when the observable modes exit the horizon. This makes it natural to treat the oscillatory part as a perturbation and expand the field as $\phi=\phi_0+\phi_1+\cdots$. Here $\phi_0$ is the solution in the absence of modulations, $\phi_1$ is linear in the modulations, and so on. We will assume that $\phi_0$ is well approximated by the slow-roll result
\be\label{eq:asrphi0}
\dot\phi_0=-\frac{V_0'(\phi_0)}{\sqrt{3 V_0(\phi_0)}}\,.
\ee
It is convenient to write the equation of motion for $\phi_1$ using $\phi_0$ as independent variable instead of time. To leading order in the potential slow-roll parameters $\epsilon_{V_0}$ and $\eta_{V_0}$ (defined in equation~\eqref{eq:Vsr}), this equation can be written as\footnote{This is only true provided $|\phi_1/f|\ll1$, but we shall see that this is the case.}
\be\label{eq:p1eom}
\phi_1''-\frac{3}{\sqrt{2\epsilon_{V_0}}}\phi_1'-\left(\frac32-\frac{3\eta_{V_0}}{2\epsilon_{V_0}}\right)\phi_1=\frac{3\Lambda^4}{2\epsilon_{V_0} f V_0(\phi_0)}\sin\left(\frac{\phi_0}{f}\right)\,,
\ee
where the slow-roll parameters are functions of $\phi_0$ and the prime indicates derivatives with respect to $\phi_0$.
Provided we are interested in a motion of $\phi_0$ that is large compared to the decay constant $f$ but small compared to the value of the inflaton when the modes that we observe in the CMB exit the horizon, we can replace $\phi_0$ by $\phi_*$ everywhere except in the argument of the sine. To leading order in slow-roll parameters and assuming $f\ll\sqrt{2\epsilon_*}$, where we use the notation $\epsilon_*\equiv\epsilon_{V_0}(\phi_*)$, the result of the equation obtained making the above substitution is
\be
\phi_1(t)=-\frac{3\Lambda^4 f}{2\epsilon_*V_0(\phi_*)}\sin\left(\frac{\phi_0(t)}{f}\right)\,.
\ee
In analogy to the linear case~\eqref{V}, it may be convenient to introduce a parameter that measures the monotonicity of the potential near the pivot scale. We define it as
\be
b_*\equiv\frac{\Lambda^4}{V_0'(\phi_*)f}\,,
\ee
where the asterisk indicates that, for all potentials except the linear potential, it depends on the pivot scale. 
Using the definition of the potential slow-roll parameters~\eqref{eq:Vsr}, the solution for the scalar field to linear order in the oscillations can then be written as\footnote{Notice that $|\phi_1/f|\lesssim\frac{3 b_* f}{\sqrt{2\epsilon_{V_0}(\phi_*    )}}\ll1$ as was needed for our derivation of equation~\eqref{eq:p1eom}.}
\be
\phi(t)=\phi_0(t)-\frac{3 b_* f^2}{\sqrt{2\epsilon_*}}\sin\left(\frac{\phi_0(t)}{f}\right)\,,
\ee
with $\phi_0(t)$ obtained from integration of equation~\eqref{eq:asrphi0}.

With this solution for the background field, it is now straightforward to calculate the Hubble slow-roll parameters~\eqref{eq:Hsr} to leading order in the oscillations. To do this, it is helpful to remember the exact relation $\dot{H}=-\dot\phi^2/2$. One finds that to leading order in the slow-roll parameters $\epsilon_{V_0}$ and $\eta_{V_0}$ as well as to leading order in the oscillations, the Hubble slow-roll parameters~\eqref{eq:Hsr} are
\be
&&\epsilon=\epsilon_*-3b_*f\sqrt{2\epsilon_*}\cos\left(\frac{\phi_0(t)}{f}\right)\,,\\
&&\delta=\epsilon_*-\eta_*-3b_*\sin\left(\frac{\phi_0(t)}{f}\right)\,.
\ee 
where $\eta_*=\eta_{V_0}(\phi_*)$\,.
From these, one obtains
\begin{equation}
\frac{\dot\delta_1}{H}=\frac{3b_*\sqrt{2\epsilon_*}}{f}\cos\left(\frac{\phi_0(t)}{f}\right)\,.
\end{equation}

As long as we are interested only in small neighborhoods of $\phi_*$, these results can be immediately translated to potentials with multiplicative corrections of the form
\be\label{eq:Vmult}
V(\phi)=V_0(\phi)\left[1+\lambda\cos\left(\frac{\phi(t)}{f}\right)\right]\,,
\ee
by replacing $\Lambda^4\to\lambda V_0(\phi_*)$, or equivalently $b_*f\sqrt{2\epsilon_*}\to\lambda$. One finds that the solution for the background scalar field is given by
\be
\phi(t)=\phi_0(t)-\frac{3 \lambda f}{2\epsilon_*}\sin\left(\frac{\phi_0(t)}{f}\right)\,,
\ee
leading to the slow-roll parameters
\be
&&\epsilon=\epsilon_*-3\lambda\cos\left(\frac{\phi_0(t)}{f}\right)\,,\\
&&\delta=\epsilon_*-\eta_*-\frac{3\lambda}{f\sqrt{2\epsilon_*}}\sin\left(\frac{\phi_0(t)}{f}\right)\,,
\ee
and finally
\begin{equation}
\frac{\dot\delta_1}{H}=\frac{3\lambda}{f^2}\cos\left(\frac{\phi_0(t)}{f}\right)\,.
\end{equation}

What remains is to find the dependence of $\phi_0$ on $X=-K\tau$, or equivalently on the conformal time which should be obtained upon integration of its equation of motion.
In the approximation we have been using in the derivation of the solution for the background scalar field, {\it i.e.} neglecting the time dependence of $\epsilon$ and working to leading order in slow-roll parameters, the relation is very simple. The equation of motion then takes the form
\be
\frac{d\phi_0}{d\ln(-\tau)}=\sqrt{2\epsilon_*}\,.
\ee 
This can immediately be integrated, and with appropriate choice of initial conditions, one finds
\be
\phi_0(\tau)=\phi_*+\sqrt{2\epsilon_*}\ln\tau/\tau_*\,.
\ee
Here $\tau_*$ is the conformal time at which the pivot scale exits the horizon and $\phi_*$ and $\epsilon_*$ are the values of the scalar field and of the slow-roll parameter at that time.
Multiplying numerator and denominator inside the logarithm by $k_* K$, making use of $-k_*\tau_*=1$, and using $X=-K\tau$, this can be written as
\be
\phi_0(X)=\phi_K+\sqrt{2\epsilon_*}\ln X\,,\;\;\;\;\text{with}\;\;\;\;\phi_K=\phi_*-\sqrt{2\epsilon_*}\ln K/k_*\,,
\ee
where $\phi_K$ is the value of the field when the mode with comoving momentum $K$ exits the horizon.
We conclude that in this approximation
\begin{equation}
\frac{\dot\delta_1}{H}=\frac{3b_*\sqrt{2\epsilon_*}}{f}\cos\left(\frac{\phi_K}{f}+\frac{\sqrt{2\epsilon_*}}{f}\ln X\right)\,.
\end{equation}

%%%%%%%%%%%%%%%%%%%%%%%%%%%%%%%%%%%%%%%%%%%%%%%%%%%%%%%%%%%%%%%%%%%%%%%%%%%%%%%%%%%%%%%%%%%%%%%%%%%%%%%%%%%%%

\section{Slow-roll Parameters}\label{a:sr}

As a courtesy to the reader, in this appendix we summarize various definitions for the choice of slow-roll parameters and provide the exact ({\it i.e.}~valid beyond the slow-roll approximation) relations between them. The potential slow-roll parameters are defined by
\be\label{eq:Vsr}
\epsilon_V&\equiv& \frac12 \left(\frac{V'}{V}\right)^2  \,, \quad \eta_V\equiv\frac{V''}{V}\,.
\ee
They are very easy to calculate because one does not need to solve for the actual dynamics of the system in order to obtain them. The smallness of $\epsilon_V$ and $\eta_V$ tells us that there exists a regime (which is often an attractor) in which the system evolves slowly. On the other hand the potential slow-roll parameters do not carry any informations about the actual dynamics (for example the inflaton could be moving fast over a flat region of the potential). They are thus of little help in order to assess the validity of the slow-roll approximation for a certain dynamics. More useful parameters are those that describe the time evolution, often referred to as Hubble slow-roll parameters. While the first slow-roll parameter is always taken to be
\be
\epsilon_{H}\equiv-\frac{\dot H}{H^2}\,,
\ee
for the second there are several different conventions:
\be
 \eta_{H}&\equiv& \frac{\dot \epsilon_{H}}{\epsilon H}=\frac{d\ln \epsilon}{d\ln a} \,,\\
  \eta_{\dot \phi}&\equiv&\frac{\ddot \phi}{\dot \phi H} =\frac{d\ln \dot \phi}{d\ln a} \,,\\
  \delta&\equiv& \frac{\ddot H}{2 H \dot H} =\frac12\frac{d\ln \dot H}{d\ln a} \,,
\ee
where $a$ is the scale factor and a dot denotes a time derivative. Notice that the slow-roll parameters are always dimensionless. A third slow-roll parameter is sometimes introduced
\be
   \xi&\equiv&\frac{{ \dddot{\phi}}}{H^2 \dot \phi}
\ee
and it is useful in the discussion of perturbations.
Exact relations to convert the various slow-roll parameters are
\be
   \epsilon_V&=&\epsilon \left(\frac{3-\eta_{\dot \phi}}{3-\epsilon}\right)^2=\epsilon \left(\frac{3+\eta_{H}/2-\epsilon}{3-\epsilon}\right)^2\,,\\
   \eta_V&=&\frac{3(\epsilon +\eta_{\dot \phi})-\xi}{3-\epsilon}=\frac{6\epsilon -\frac32 \eta_{H}-\xi}{3-\epsilon}\\
   \eta_{\dot \phi}&=&\epsilon-\frac{\eta_H}{2}=-\delta\,,\\
   \eta_H&=&2\delta +2\epsilon\,.
\ee

%%%%%%%%%%%%%%%%%%%%%%%%%%%%%%%%%%%%%%%%%%%%%%%%%%%%%%%%%%%%%%%%%%%%%%%%%%%%%%%%%%%%%%%%%%%%%%%%%%%%%%%%%%%%%

\end{document}